\documentclass[sigconf]{acmart}
\usepackage{amsmath}
\usepackage{setspace}%行间距
\usepackage{geometry}
\usepackage{subfigure}
\usepackage[inkscapelatex=false]{svg}
\usepackage[ruled]{algorithm2e}
\usepackage{enumitem}

\newcommand{\ie}[0]{\textit{i.e.}}
\newcommand{\eg}[0]{\textit{e.g.}}

\usepackage{multirow}
\AtBeginDocument{%
  \providecommand\BibTeX{{%
    \normalfont B\kern-0.5em{\scshape i\kern-0.25em b}\kern-0.8em\TeX}}}

\copyrightyear{2023}
\acmYear{2023}
\setcopyright{rightsretained}
\acmConference[SIGIR '23]{Proceedings of the 46th International ACM SIGIR Conference on Research and Development in Information Retrieval}{July 23--27, 2023}{Taipei, Taiwan}
\acmBooktitle{Proceedings of the 46th International ACM SIGIR Conference on Research and Development in Information Retrieval (SIGIR '23), July 23--27, 2023, Taipei, Taiwan}\acmDOI{10.1145/3539618.3591734}
\acmISBN{978-1-4503-9408-6/23/07}

\begin{document}

\title{Multi-behavior Self-supervised Learning for Recommendation}

\author{Jingcao Xu}
\email{xjc20@mails.tsinghua.edu.cn}
\affiliation{%
  \institution{Tsinghua University}
  % \streetaddress{P.O. Box 1212}
  \city{Beijing}
  \country{China}
}
\author{Chaokun Wang}
\authornote{Chaokun Wang is a corresponding author.}
\email{chaokun@tsinghua.edu.cn}
\affiliation{%
  \institution{Tsinghua University}
  \city{Beijing}
  \country{China}
}

\author{Cheng Wu}
\email{c-wu19@mails.tsinghua.edu.cn}
\affiliation{%
  \institution{Tsinghua University}
  \city{Beijing}
  \country{China}
}
\author{Yang Song}
\email{yangsong@kuaishou.com}
\affiliation{%
  \institution{Kuaishou Inc.}
  \city{Beijing}
  \country{China}
}
\author{Kai Zheng}
\email{zhengkai@kuaishou.com}
\affiliation{%
  \institution{Kuaishou Inc.}
  \city{Beijing}
  \country{China}
}
\author{Xiaowei Wang}
\email{wangxiaowei03@kuaishou.com}
\affiliation{%
  \institution{Kuaishou Inc.}
  \city{Beijing}
  \country{China}
}
\author{Changping Wang}
\email{wcpvincent@gmail.com}
\affiliation{%
  \institution{Kuaishou Inc.}
  \city{Beijing}
  \country{China}
}
\author{Guorui Zhou}
\email{zhouguorui@kuaishou.com}
\affiliation{%
  \institution{Kuaishou Inc.}
  \city{Beijing}
  \country{China}
}
\author{Kun Gai}
\email{gai.kun@qq.com}
\affiliation{%
  \institution{Unaffiliated}
  \city{Beijing}
  \country{China}
}

\renewcommand{\shortauthors}{Jingcao Xu et al.}

\begin{abstract}
Modern recommender systems often deal with a variety of user interactions, \eg, click, forward, purchase, etc., which requires the underlying recommender engines to fully understand and leverage multi-behavior data from users.
Despite recent efforts towards making use of heterogeneous data, multi-behavior recommendation still faces great challenges.
Firstly, sparse target signals and noisy auxiliary interactions remain an issue. Secondly, existing methods utilizing self-supervised learning (SSL) to tackle the data sparsity neglect the serious optimization imbalance between the SSL task and the target task.
% \todo{i) Increasing the robustness to sparse supervision signal and interaction noises. ii) Mitigating the optimization imbalance between auxiliary tasks and the target task.}
% To tackle the above challenges, we propose a
% Multi-Behavior Self-Supervised Learning (MBSSL) framework that conducts self-supervised learning at two different levels and  modifies the direction and magnitude of gradients adaptively.
Hence, we propose a
Multi-Behavior Self-Supervised Learning (MBSSL) framework together with an adaptive optimization method.
Specifically, we devise a behavior-aware graph neural network incorporating the self-attention mechanism to capture behavior multiplicity and dependencies. To increase the robustness to data sparsity under the target behavior and noisy interactions from auxiliary behaviors, we propose a novel self-supervised learning paradigm to conduct node self-discrimination at both inter-behavior and intra-behavior levels. In addition, we develop a customized optimization strategy through hybrid manipulation on gradients to adaptively balance the self-supervised learning task and the main supervised recommendation task. Extensive experiments on five real-world datasets demonstrate the consistent improvements obtained by MBSSL over ten state-of-the-art (SOTA) baselines. We release our model implementation at: https://github.com/Scofield666/MBSSL.git.
\end{abstract}





\maketitle

% \enlargethispage{2em}
\vspace{-0.25cm}
\section{Introduction}
Recommender systems have emerged as indispensable means of promoting personalized suggestions for users in a variety of applications, ranging from e-commerce platforms~\cite{gu2020hierarchical}, online video websites~\cite{chang2021sequential} and location-based services ~\cite{altenburger2019yelp}.
Collaborative Filtering (CF) is the most extensively adopted paradigm for recommendation, which develops from traditional Matrix Factorization algorithms ~\cite{he2016fast} to novel Neural Network (NN) architectures like Autoencoders ~\cite{sedhain2015autorec} or Graph Neural Networks (GNNs) ~\cite{wang2019neural, he2020lightgcn}.

% Recent advances in deep learning models have inspired researchers to augment CF architecture with Neural Networks (NNs) to capture more complex latent semantics of user-item correlations.
% For instance, NCF~\cite{he2017neural} employs Multi-Layer Perceptron (MLP) as the interaction function, whereas AutoRec~\cite{sedhain2015autorec}, CDAE~\cite{wu2016collaborative}, Multi-VAE and Multi-DAE ~\cite{liang2018variational} utilize specific auto-encoder variants for modeling.
% Further, Graph Neural Networks (GNNs) have sprung up for their capability of preserving high-order structural information, where the representations are learned via iteratively passing and aggregating the neighborhood message layer by layer~\cite{wang2019neural, he2020lightgcn}.

% Nevertheless, these CF solutions are mostly designed based on singular-type of user-item interaction behavior, whereas the real-world recommendation is more like a multi-behavior setting, \ie, users’ behaviors are multi-typed which involve heterogeneous relations between users and items ~\cite{guo2019buying, xia2021knowledge, cen2019representation, xia2020multiplex}. For example, in the online video platform, users can interact with videos in multiple manners, including view, like, comment, and so on.
% \enlargethispage{2em}
Nevertheless, most CF solutions are designed based on a singular-type of user-item interaction behavior, whereas the real-world recommendation is more like a multi-behavior setting, \ie, the interaction behaviors between users and items are multi-typed with one target behavior to be optimized ~\cite{xia2021knowledge, xia2020multiplex}.
% For example, in the online video platform, users can interact with videos in multiple manners, including view, like, comment, and so on.
For example, in E-commerce services, users can interact with items in multiple manners, including page view, favorite and purchase, among which purchase behavior is the optimization target as it is directly related to Gross Merchandise Value (GMV).

\begin{figure*}[ht]
\centering
    \setlength{\belowcaptionskip}{-1em}
    \includegraphics[width=0.95\linewidth]{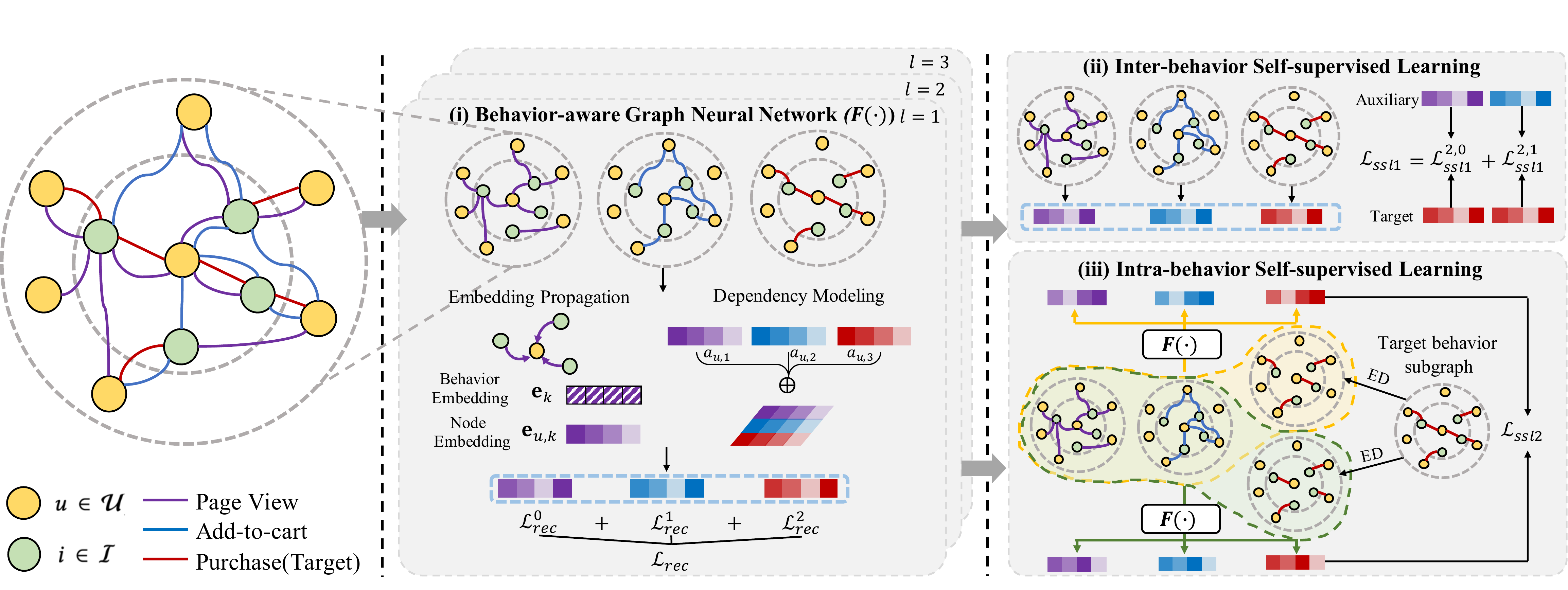}
    \caption{The overall architecture of MBSSL. i) Behavior-aware graph neural network ($\boldsymbol{F(\cdot)}$) which performs behavior-specific embedding propagation and cross-behavior dependency modeling. ii) Inter-behavior SSL contrasting between target and auxiliary subgraphs to alleviate skewed data distribution and data sparsity under the target behavior. iii) Intra-behavior SSL contrasting between target and Edge Dropout (ED) subgraphs to counteract interaction noises from auxiliary behaviors.
    }
    \label{fig:intro_example}
\end{figure*}

Up to now, great efforts on multi-behavior recommendation have been made towards modeling the complex semantics of different interaction behaviors. MATN ~\cite{xia2020multiplex} and MBGMN ~\cite{xia2021graph} encode the interaction pattern multiplicity and behavior dependencies via memory-augmented Transformer and Graph Meta Network respectively. To characterize discriminative semantics of different behaviors, some research works ~\cite{chen2020efficient, chen2021graph} learn one specific representation for each behavior. In addition, one recent study CML ~\cite{wei2022contrastive} firstly incorporates contrastive learning to tackle the label scarcity problem for the target behavior. Despite their success,
% Different types of behavior exhibit discriminative semantic information for characterizing users’ personalized interests. Hence, encoding the interaction pattern multiplicity and exploring the behavior dependencies are extremely significant. However, it is non-trivial to leverage heterogeneous types of data in assisting the recommendation.
the multi-behavior recommendation still encounters the following challenges.

% \noindent \textbf{Model behavior multiplicity and dependencies.} On one hand, each type of interaction behavior presents distinct characteristics in nature. For example, in the online video platform, the click behavior appears more frequently than the like behavior while the latter contributes more to reflecting user preferences. On the other hand, different types of interaction behavior may be correlated in complex ways and the pair-wise dependencies may vary by user, based on the personalized interests. Thus, the obtained representations are expected to embrace behavior-specific semantics and capture the implicit correlations between behaviors.

% \enlargethispage{2em}
\noindent \textbf{The robustness against data sparsity and interaction noises.} In spite of the fact that interaction data of auxiliary behaviors could offer quite complementary information for recommendation on the target behavior, the data sparsity under the target behavior remains an issue.
One possible solution is proposed in CML ~\cite{wei2022contrastive} which makes full use of supervision signals from auxiliary behaviors via conducting contrastive learning between each auxiliary behavior and target behavior pair. However, auxiliary behaviors might contain noisy interactions which are detrimental to the target task at the same time. Hence, simply adopting the contrastive learning paradigm in CML is likely to exacerbate the negative transfer towards noise distributions in auxiliary behaviors, greatly undermining the true semantics of the target behavior.
In this regard, an approach to making comprehensive and adaptive use of interaction data plays a vital role in performance enhancement.
% Specifically, the learned representations under sparse data suffer from great deviation; the representations under different behaviors exhibit bias disparity resulting from the skewed data distribution across different types of interaction behavior, which undermines the recommendation performance greatly.

% \enlargethispage{2em}
\noindent \textbf{The optimization imbalance between auxiliary tasks and the target task.} Existing multi-behavior recommendation solutions basically adopt the Multi-task Learning (MTL) paradigm to optimize the auxiliary and target task jointly~\cite{gao2019neural, xia2020multiplex, jin2020multi, chen2020efficient, chen2021graph}. However, ignoring the estimation of contributions of each task to the optimization target will suffer from a serious optimization imbalance problem where auxiliary tasks might dominate the network weights, resulting in worse performance on the target task. What’s more, existing multi-task learning methods like GradNorm ~\cite{chen2018gradnorm} or PCGrad~\cite{yu2020gradient} do not apply to the scenario where the self-supervised learning (SSL) task is treated as the auxiliary task, in that the SSL task has a confounding effect on the target task, depending on the particular design of SSL. Therefore, another key problem in multi-behavior recommendation is the elaborated design of the optimization method to mitigate the optimization imbalance between the auxiliary and target tasks.

Towards this end, we propose a \textbf{M}ulti-\textbf{B}ehavior \textbf{S}elf-\textbf{S}upervised \textbf{L}earning solution, short-handed as \textbf{MBSSL}, to ensure the performance for multi-behavior recommendation.

Specifically, we first devise a behavior-aware graph neural network augmented with behavior representation learning and the self-attention mechanism to jointly model the behavior inner-context and behavior inter-dependencies.
% In our graph neural network, we inject behavior context into the message passing architecture to learn contextualized representations, and model cross-behavior dependencies via the self-attention mechanism to further enrich the representations.
For dealing with the sparse supervision signals under the target behavior, a comprehensive self-supervised learning paradigm is introduced to contrast nodes from inter-behavior and intra-behavior levels respectively.
% The inter-behavior SSL \zktodo{following CML \cite{wei2022contrastive}} supplements the target behavior with informative semantics from auxiliary behaviors as well as alleviates the bias gap of learned representations under different behaviors.
The inter-behavior SSL transfers informative semantics from auxiliary behaviors to the target behavior via selectively constructing negative node pairs.
To further increase the robustness to noisy interactions, the intra-behavior SSL consolidates the self-supervised information in the target behavior to  counteract the potential negative transfer brought by inter-behavior SSL.
In addition, based on the observation that the SSL task exhibits an arbitrary optimization trend with respect to the target task, we design a multi-behavior optimization method that hybridly rectifies the direction and magnitude of gradients to balance SSL tasks and the target task in optimization.

In a nutshell, our main contributions are as follows:\\
$\bullet$\, We develop a new self-supervised learning framework named MBSSL for multi-behavior recommendation,
% which uncovers latent cross-behavior dependencies as well as alleviates data sparsity issues under the target behavior.
% Besides, we are the first to research the optimization imbalance problem between the SSL task and the target recommendation task with a proposed hybrid gradient manipulation method (Section~\ref{sec:method_adaptive}).
which embodies a behavior-aware graph neural network (Section \ref{sec:method_bgnn}) to uncover latent cross-behavior dependencies and a comprehensive SSL paradigm (Section \ref{sec:method_hssl}) at inter-behavior and intra-behavior levels to alleviate the problem of data sparsity and interaction noises. \\
% \item In MBSSL, the inter-behavior SSL (Section \ref{sec:method_inter}) contrasts between the auxiliary behavior and target behavior while the intra-behavior SSL (Section \ref{sec:method_intra}) contrasts between augmented views originating from the target behavior. Such two levels of SSL enable to distill of auxiliary supervision signals and figure out informative graph patterns respectively.
$\bullet$\, To the best of our knowledge, we are the first to research the optimization imbalance problem between the SSL task and the target recommendation task. Accordingly, we propose a hybrid gradient manipulation method on both the magnitude and direction to adjust the optimization trend (Section ~\ref{sec:method_adaptive}).\\
$\bullet$\, Extensive experiments are conducted on five real-world datasets compared with ten SOTA baselines (Section ~\ref{sec:exp}). The consistent performance uplift on two representative metrics demonstrates the effectiveness of MBSSL.
% \end{itemize}
% \zktodo{In above summary, maybe put the contribution of 3.3 in a separate paragraph is better, or put 3.1 and 3.2 in a paragraph, 3.3 in another paragraph, if space is limited.}

% \enlargethispage{2em}

% \enlargethispage{2em}
% \vspace{-0.3cm}
\section{Preliminary}
\label{sec:preliminary}
In typical recommender systems, we define the set of users and items as $\mathcal{U}$ ($u \in \mathcal{U}$) and $\mathcal{I}$ ($i \in \mathcal{I}$) respectively, where $|\mathcal{U}|$ and $|\mathcal{I}|$ represent the number of users and items. $\mathcal{G} = (\mathcal{V}, \mathcal{E})$ is a bipartite graph based on the user-item interactions, where $\mathcal{V} = \mathcal{U} \cup \mathcal{I}$ is denoted as the whole node set involving users and items, and the edge set $\mathcal{E}$ represents the observed interactions. In the multi-behavior scenario with $K$ types of behaviors, the whole bipartite graph $\mathcal{G}$ can be segmented into $K$ behavior subgraphs $\{\mathcal{G}_1, \mathcal{G}_2, .., \mathcal{G}_K\}$ based on the interaction behavior type. To reflect the multi-behavior interaction data, we define a tensor $\mathbf{X} \in \mathbb{R}^{|\mathcal{U}| \times |\mathcal{I}| \times K}$, where the individual element $x_{u,i}^k \in \mathbf{X}$ equals to 1 if $u$ interacted with $i$ under behavior $k$, otherwise $x_{u,i}^k=0$. Therefore, the task of multi-behavior recommendation is formulated as:

\textbf{Input}: The multi-behavior interaction tensor $\mathbf{X} \in \mathbb{R}^{|\mathcal{U}| \times |\mathcal{I}| \times K}$ under multiplex $K$ types of behaviors.

\textbf{Output}: A recommendation model that estimates the probability that a user $u$ interacts with item $i$ under the $K$-th behavior, \ie, the target behavior.

\vspace{-0.2cm}
\section{Methodology}
\label{sec:method}
In this section, our MBSSL framework is presented in detail.
The architecture of MBSSL is depicted in Figure 1, which encapsulates three key components:
i) behavior-aware graph neural network which collectively captures behavioral context and dependencies (Section \ref{sec:method_bgnn}),
% ii) hierarchical self-supervised learning, which increases the robustness to data sparsity and noisy interactions by conducting inter-behavior and intra-behavior self-supervised learning respectively.
ii) inter-behavior self-supervised learning to facilitate knowledge transfer, and
iii) intra-behavior self-supervised learning to counteract noisy interactions.
% Also, we propose the adaptive multi-behavior optimization for balancing SSL tasks and the target task.
% % \todo{iii) adaptive multi-task learning optimization, which }
% Finally, we discuss the differences between existing models and ours (Section \ref{sec:method_comp}).

% In this section, we introduce our proposed MBSSL framework in detail and discuss the differences from existing models (Section \ref{sec:method_comp}). The overall architecture of MBSSL is depicted in Figure 1, which encapsulates three key components:
% i) behavior-aware graph neural network which collectively captures behavioral context and dependencies (Section \ref{sec:method_bgnn}),
% % ii) hierarchical self-supervised learning, which increases the robustness to data sparsity and noisy interactions by conducting inter-behavior and intra-behavior self-supervised learning respectively.
% ii) inter-behavior self-supervised learning to facilitate knowledge transfer (Section \ref{sec:method_inter}),
% iii) intra-behavior self-supervised learning to counteract noisy interactions (Section \ref{sec:method_intra}).
% Besides, we propose the adaptive multi-behavior optimization for balancing SSL tasks and the target task in Section \ref{sec:method_adaptive}.
% % \todo{iii) adaptive multi-task learning optimization, which }

\enlargethispage{2em}
\vspace{-0.2cm}
\subsection{Behavior-aware Graph Neural Network}
\label{sec:method_bgnn}
Motivated by the message-passing architecture in GNNs, we develop a behavior-aware graph neural network to capture complex CF signals between nodes and among behaviors. More concretely, we first obtain contextualized node embeddings under each behavior subgraph via behavior-specific embedding propagation. Thereafter, we enhance the embeddings with personalized behavioral correlations via cross-behavior dependency modeling.
\vspace{-0.2cm}
\subsubsection{Behavior-specific Embedding propagation}
Under such a multi-behavior setting, we first construct each bipartite behavior subgraph $\mathcal{G}_k$ according to the interaction behavior type $k$ and then we perform embedding propagation on each subgraph to obtain the representation of each node under each behavior. In order to explicitly manifest the discriminative semantic of each behavior and capture the contextualized user preference, we also embed behaviors on top of nodes ~\cite{schlichtkrull2018modeling, chen2020efficient, chen2021graph} and incorporate the representations of each behavior into the message-passing paradigm as:
% \begin{align}
%     \boldsymbol{\rm e}_{u,k}^{(l+1)} = \textbf{AGG}(\{\boldsymbol{\rm e}_{i,k}^{(l)}: i \in \mathcal{N}_{u,k}\}, \boldsymbol{\rm e}_k^{(l)}),
% \end{align}

\vspace{-0.3cm}
\begin{spacing}{1}
\begin{equation}
    \begin{aligned}
        \boldsymbol{\rm e}_{u,k}^{(l+1)} = {\rm LeakyRelu}(\boldsymbol{\rm W}^{(l)} \cdot {\rm mean}(\{{\boldsymbol{\rm e}_{i,k}^{(l)} \odot \boldsymbol{\rm e}_k^{(l)}: i \in \mathcal{N}_{u,k}\})}),
    \end{aligned}
\end{equation}
\end{spacing}

% \begin{align}
%     \boldsymbol{\rm e}_{u,k}^{(l+1)} = {\rm LeakyRelu}(\boldsymbol{\rm W}^{(l)} \cdot {\rm mean}(\{{\boldsymbol{\rm e}_{i,k}^{(l)} \odot \boldsymbol{\rm e}_k^{(l)}: i \in \mathcal{N}_{u,k}\})}),
% \end{align}

\noindent where $\boldsymbol{\rm e}_{u,k}^{(l+1)}$ denotes the embedding of node $u$ under behavior $k$ in the $(l+1)$-th propagation layer; $\mathcal{N}_{u,k}$ is the set of immediate neighbors of $u$ under behavior type $k$; $\boldsymbol{\rm W}^{(l)}$ is layer-specific and $\odot$ denotes the element-wise product of two vectors; $\boldsymbol{\rm e}_{k}^{(l)}$ represents the embedding of behavior $k$ in the $l$-th layer, which is updated by multiplying another layer-specific parameter $\boldsymbol{\rm W}_{b}$:
\vspace{-0.3cm}
\begin{spacing}{1}
\begin{equation}
    \begin{aligned}
    \boldsymbol{\rm e}_{k}^{(l+1)} = \boldsymbol{\rm W}_{b}^{(l)} \boldsymbol{\rm e}_{k} ^ {(l)}.
    \end{aligned}
\end{equation}
\end{spacing}

% \begin{align}
%     \boldsymbol{\rm e}_{k}^{(l+1)} = \boldsymbol{\rm W}_{b}^{(l)} \boldsymbol{\rm e}_{k} ^ {(l)}.
% \end{align}

% For accurately modeling the user preference under the specific context, we propose to fuse the behavior embedding $\boldsymbol{\rm e}_k$ and neighbor's embedding $\boldsymbol{\rm e}_{i,k}$ during the propagation process. Hence, the \textbf{AGG} function is defined as:
% \begin{align}
%         \boldsymbol{\rm e}_{u,k}^{(l+1)} = {\rm LeakyRelu}(\boldsymbol{\rm W}^{(l)} \cdot {\rm mean}(\{{\boldsymbol{\rm e}_{i,k}^{(l)} \odot \boldsymbol{\rm e}_k^{(l)}: i \in \mathcal{N}_{u,k}\})}),
% \end{align}
% where $\boldsymbol{\rm W}^{(l)}$ is another layer-specific parameter and $\odot$ denotes the element-wise product of two vectors.
% The basic idea of widely adopted message-passing architecture is to propagate the aggregate the neighborhood information iteratively layer by layer
\vspace{-1em}
\subsubsection{Cross-behavior dependency modeling} Given that different behaviors would interweave with each other in an implicit manner and the correlations among behaviors vary by user, we leverage the self-attention~\cite{lin2017structured} mechanism to model the cross-behavior dependency.

Specifically, we concatenate the embeddings of node $u$ under all the behaviors as $\boldsymbol{e}_u \in \mathbb{R}^{K \times d}$, and then the coefficients $\boldsymbol{\rm a}_{u,k} \in \mathbb{R}^K $ reflecting the dependency between behavior $k$ and other behaviors for user $u$ are computed by:
\vspace{-0.3cm}
\begin{spacing}{1}
\begin{equation}
    \begin{aligned}
      \boldsymbol{\rm a}_{u,k} = {\rm softmax} ((\boldsymbol{\rm W}_2^k)^T {\rm tanh} ((\boldsymbol{e}_u \boldsymbol{\rm W}_1^k)^T)),
    \end{aligned}
\end{equation}
\end{spacing}
% \begin{align}
%     \boldsymbol{\rm a}_{u,k} = {\rm softmax} ((\boldsymbol{\rm W}_2^k)^T {\rm tanh} ((\boldsymbol{e}_u \boldsymbol{\rm W}_1^k)^T)),
% \end{align}
\noindent where $\boldsymbol{\rm W}_1^k \in \mathbb{R}^{d \times d^{\prime}}, \boldsymbol{\rm W}_2^k \in \mathbb{R}^{d^{\prime}}$ are two behavior-specific parameters and $d ^ {\prime}$ is the output dimension size. Hence, the enhanced embedding of node $u$ under behavior $k$ can be readily calculated as:
\vspace{-0.6cm}
\begin{spacing}{1}
\begin{equation}
    \begin{aligned}
      \boldsymbol{e}_{u,k} = \boldsymbol{\rm a}_{u,k} \boldsymbol{e}_u.
    \end{aligned}
\end{equation}
\end{spacing}
% \begin{align}
%     \boldsymbol{e}_{u,k} = \boldsymbol{\rm a}_{u,k} \boldsymbol{e}_u.
% \end{align}

Since the embeddings of different layers express different connections, we utilize the mean-pooling to integrate the embeddings of all layers as follows:
\vspace{-0.3cm}
\begin{spacing}{1}
\begin{equation}
    \begin{aligned}
      \boldsymbol{\rm e}_{u,k} = \frac{1}{L}\sum_{l=0}^{L-1} \boldsymbol{\rm e}_{u,k}^{(l)}; \quad
    \boldsymbol{\rm e}_{k} = \frac{1}{L}\sum_{l=0}^{L-1} \boldsymbol{\rm e}_{k}^{(l)};
    \end{aligned}
\end{equation}
\end{spacing}

% \begin{align}
%     \boldsymbol{\rm e}_{u,k} = \frac{1}{L}\sum_{l=0}^{L-1} \boldsymbol{\rm e}_{u,k}^{(l)}; \quad
%     \boldsymbol{\rm e}_{k} = \frac{1}{L}\sum_{l=0}^{L-1} \boldsymbol{\rm e}_{k}^{(l)};
% \end{align}
\enlargethispage{2em}
\vspace{-0.4cm}
\subsection{Multi-behavior Self-supervised Learning}
\label{sec:method_hssl}
As mentioned before, sparse supervision signals of the target behavior may lead to severe bias of learned representations compared with those of auxiliary behaviors. Besides, the overlook of noisy interactions brought from auxiliary behaviors would exaggerate the immoderate reliance on certain interactions. Accordingly, we introduce a novel self-supervised learning paradigm to conduct self-discrimination contrastive learning from both inter-behavior and intra-behavior levels.
\vspace{-0.2cm}
\subsubsection{Inter-behavior Self-supervised Learning}
\label{sec:method_inter}
In view of the fact that supervision signals in auxiliary behaviors are much richer than that in the target behavior, we perform selective contrastive learning between auxiliary behaviors and the target behavior to enable knowledge transfer, and thus alleviate the data sparsity at the first step.

% Intuitively, the interaction density of different behaviors exhibits such great discrepancy that the learned node representations under each behavior struggle to coincide for consistently revealing the preference of one specific user. Thus, we propose to perform contrastive learning between the target behavior and the auxiliary behaviors, in which way the semantics of auxiliary behaviors could be transferred and the skewed data distribution of different behaviors could be alleviated to some extent.

In particular, each auxiliary behavior $k$ from $\{1, 2, ..., K-1\}$ is to be contrasted with the target behavior $K$ to provide distinct semantics.
% Different from existing SSL solutions ~\cite{hafidi2020graphcl, zhu2021graph} which generate and contrast between augmented views, we directly regard the auxiliary behavior subgraph $\mathcal{G}_k$ as the augmented view of target behavior subgraph $\mathcal{G}_K$.
The common practice will treat the views of the same node as positive pairs (\ie, $\{(\boldsymbol{\rm e}_{u,K}, \boldsymbol{\rm e}_{u,k}) | u \in \mathcal{U}\}$), and the views of any different nodes as the negative pairs  (\ie, $\{(\boldsymbol{\rm e}_{u,K}, \boldsymbol{\rm e}_{v,k}) | u, v \in \mathcal{U}, u \neq v \}$).
% The positive pair offers auxiliary supervision signals to the target behavior and narrows the skewed data distribution gap while the negative pair guarantees the discriminative characteristics of each node.
However, in the recommendation setting which encapsulates various interactions, the same two subjects will have some commonalities (\eg, users share similar preferences or items have similar attributes.). In this case, the constructed negative pairs following the common practice are likely to include many false negatives (\ie, highly similar nodes), which will discard true semantic information ~\cite{huynh2022boosting}. Therefore, we propose to find potential false negatives based on the calculated similarity score using swing algorithm ~\cite{yang2020large} and eliminate them when contrasting node pairs.

To be more specific, the similarity score of two users $u, v$ in subgraph $\mathcal{G}_k (k \in \{1, 2, ..., K\}$) using swing algorithm is calculated as:
\vspace{-0.3cm}
\begin{spacing}{1}
\begin{equation}
    \begin{aligned}
    S_{\mathcal{G}_k}(u, v) = \sum_{i \in \mathcal{N}_{u,k} \cap \mathcal{N}_{v,k}} \sum_{j \in \mathcal{N}_{u,k} \cap \mathcal{N}_{v,k}} \frac{1}{\alpha +|\mathcal{N}_{i,k}\cap \mathcal{N}_{j,k}|}
    \end{aligned}
\end{equation}
\end{spacing}
\noindent where $\alpha$ is a smoothing coefficient. Then the final similarity score $S(u, v)$ is the average of the score in each subgraph $S_{\mathcal{G}_k}(u, v)$.
% \begin{spacing}{1}
% \begin{equation}
%     \begin{aligned}
%     S(u, v) = \frac{1}{K}\sum_{k=1}^K S_{\mathcal{G}_k}(u, v)
%     \end{aligned}
% \end{equation}
% \end{spacing}

Then we define the false negatives for $u$ as users with top-$N$ similarity scores, $FN(u)=\{v | S(u, v) \in top(S(u), N)\}$. Accordingly, the inter-behavior contrastive loss between the target behavior $K$ and auxiliary behavior $k$ which eliminates specific false negatives is defined as:

% Following ~\cite{chen2020simple}, we adopt InfoNCE~\cite{gutmann2010noise} loss to maximize the agreement of positive pairs and minimize that of negative pairs, then the inter-behavior contrastive loss between the target behavior $K$ and  auxiliary behavior $k$ is defined as:
\vspace{-0.3cm}
\begin{spacing}{1}
\begin{equation}
    \begin{aligned}
    \mathcal{L}_{ssl_1, user} ^ {K, k} = \sum_{u \in \mathcal{U}} -{\rm log} \frac{{\rm exp} (\phi (\boldsymbol{\rm e}_{u,K}, \boldsymbol{\rm e}_{u,k})/ \tau)}{\sum_{v \in \mathcal{U} \backslash FN(u)} {\rm exp} (\phi (\boldsymbol{\rm e}_{u,K}, \boldsymbol{\rm e}_{v,k})/ \tau)},
    \end{aligned}
\end{equation}
\end{spacing}

\noindent where $\tau$ is the temperature hyperparameter in \textit{softmax} and $\phi(\cdot)$ denotes the inner-product of two vectors. When analogously combined with the contrastive loss of the item side, the inter-behavior contrastive loss between the target behavior $K$ and auxiliary behavior $k$ is $\mathcal{L}_{ssl_1} ^ {K, k}= \mathcal{L}_{ssl_1, user} ^ {K, k} + \mathcal{L}_{ssl_1, item} ^ {K, k}$. Therefore, the ultimate inter-behavior contrastive loss is the sum of each pair of auxiliary behavior and target behavior as: $\mathcal{L}_{ssl1} =\mathcal{L}_{ssl_1} ^ {K, 1} + ... + \mathcal{L}_{ssl_1} ^ {K, k}+...+\mathcal{L}_{ssl_1} ^ {K, K-1} $.

\vspace{-0.2cm}
\subsubsection{Intra-behavior Self-supervised Learning}
\label{sec:method_intra}
% topology-based and semantic-based
% uniformity to evaluate the embeddings of SSL
For alleviating the skewed data distribution across different behaviors, the inter-behavior self-supervised learning encourages the similarity of node representations under the target behavior and auxiliary behaviors. However, in view of the higher proportion of noisy interactions under auxiliary behaviors, more noises would be implicitly transferred into the target behavior as well, making the learned representations dominated by auxiliary signals while lose the intrinsic semantics under the target behavior. Hence, we devise an intra-behavior self-supervised learning to generate and contrast structurally-augmented views of the target behavior subgraph,
in which way we consolidate and amplify the impact of supervision signals within the target behavior itself to counteract the negative transfer towards noise distributions in auxiliary behaviors.
Specifically, we first generate two augmented views $\mathcal{G}_K^1, \mathcal{G}_K^2$ from the target behavior subgraph by performing edge dropout introduced in ~\cite{wu2021self}. We denote $\mathcal{G}_K=(\mathcal{V}_K, \mathcal{E}_K)$ as the target behavior subgraph, and then two augmented views  are elaborated as:
\vspace{-0.2cm}
\begin{spacing}{1}
\begin{equation}
    \begin{aligned}
    \mathcal{G}_K^1 = (\mathcal{V}_K, \mathbf{M}_1 \odot \mathcal{E}_K), \quad
    \mathcal{G}_K^2 = (\mathcal{V}_K, \mathbf{M}_2 \odot \mathcal{E}_K),
    \end{aligned}
\end{equation}
\end{spacing}

% \begin{align}
%     \mathcal{G}_K^1 = (\mathcal{V}_K, \mathbf{M}_1 \odot \mathcal{E}_K), \quad
%     \mathcal{G}_K^2 = (\mathcal{V}_K, \mathbf{M}_2 \odot \mathcal{E}_K),
% \end{align}
\noindent where $\mathbf{M}_1, \mathbf{M}_2 \in \{0,1\}^{|\mathcal{E}_K|}$ are two random masking vectors controlling the kept edge set with the same dropout out ratio $\rho$.

After encoding the two augmented views together with auxiliary behavior subgraphs respectively, we obtain the node representations of augmented views, denoted as $\boldsymbol{\rm e}_{u,K}^1, \boldsymbol{\rm e}_{u, K}^2$. Similar to Section \ref{sec:method_inter}, we then contrast views at the node scale with positive pairs as $\{(\boldsymbol{\rm e}_{u,K}^1, \boldsymbol{\rm e}_{u,K}^2) | u \in \mathcal{U}\}$ and negative pairs as $\{(\boldsymbol{\rm e}_{u,K}^1, \boldsymbol{\rm e}_{v,K}^2) | u, v \in \mathcal{U}, u \neq v \}$. The optimization objective of user side is calculated via InfoNCE loss likewise as:
\vspace{-0.3cm}
\begin{spacing}{1}
\begin{equation}
    \begin{aligned}
    \mathcal{L}_{ssl_2, user} = \sum_{u \in \mathcal{U}} -{\rm log} \frac{{\rm exp} (\phi (\boldsymbol{\rm e}_{u,K}^1, \boldsymbol{\rm e}_{u,K}^2)/ \tau)}{\sum_{v \in \mathcal{U}} {\rm exp} (\phi (\boldsymbol{\rm e}_{u,K}^1, \boldsymbol{\rm e}_{v,K}^2)/ \tau)}.
    \end{aligned}
\end{equation}
\end{spacing}

% \begin{align}
%     \mathcal{L}_{ssl_2, user} = \sum_{u \in \mathcal{U}} -{\rm log} \frac{{\rm exp} (\phi (\boldsymbol{\rm e}_{u,K}^1, \boldsymbol{\rm e}_{u,K}^2)/ \tau)}{\sum_{v \in \mathcal{U}} {\rm exp} (\phi (\boldsymbol{\rm e}_{u,K}^1, \boldsymbol{\rm e}_{v,K}^2)/ \tau)}.
% \end{align}

Analogously combining the loss of item side, we obtain the final objective function of intra-behavior self-supervised learning as $\mathcal{L}_{ssl_2} = \mathcal{L}_{ssl_2, user} + \mathcal{L}_{ssl_2, item}$.

% \vspace{-0.3cm}
% \enlargethispage{2em}
\subsection{Adaptive Multi-behavior Optimization}
In this subsection, we present our optimization objective and a novel adaptive optimization method for multi-behavior setting.
% Finally, the analysis on the time complexity of our model is provided.
\label{sec:method_adaptive}
\vspace{-0.2cm}
\subsubsection{Optimization Objective} For learning model parameters in an effective and stable way, we leverage a recently proposed non-sampling objective for recommendation ~\cite{chen2020efficient} which has been proved to be superior to the traditional Bayesian Personalized Ranking (BPR) loss. For a specific batch of users $\mathcal{B}$ and the whole item set $\mathcal{I}$, the non-sampling recommendation loss under behavior $k$ is:

\vspace{-0.3cm}
\begin{spacing}{1}
\begin{equation}
    \begin{aligned}
    \label{s}
    \begin{split}
    \mathcal{L}_{rec}^k = \sum_{u \in \mathcal{B}} \sum_{i \in \mathcal{I}_{u}^{k+}} \left( (c_i^{k+}-c_i^{k-}) ({\hat x}_{u,i}^k)^{2} - 2c_i^{k+}{\hat x}_{u,i}^k \right) \\
    % \vspace{-1cm}
    + \sum_{m=1}^d \sum_{n=1}^d \Big(
    \big (\boldsymbol{\rm e}_{k}^{(m)} \boldsymbol{\rm e}_{k}^{(n)} \big)
    \big( \sum_{u \in \mathcal{B}} \boldsymbol{\rm e}_{u,k}^{(m)} \boldsymbol{\rm e}_{u,k}^{(n)} \big)
    \big( \sum_{i \in \mathcal{I}} \boldsymbol{\rm e}_{i,k}^{(m)} \boldsymbol{\rm e}_{i,k}^{(n)} \big)
    \Big),
\end{split}
    \end{aligned}
\end{equation}
\end{spacing}

% \begin{align}
% \label{s}
% \begin{split}
%     \mathcal{L}_{rec}^k = \sum_{u \in \mathcal{B}} \sum_{i \in \mathcal{I}_{(u)}^{k+}} \left( (c_i^{k+}-c_i^{k-}) ({\hat x}_{u,i}^k)^{2} - 2c_i^{k+}{\hat x}_{u,i}^k \right) \\
%     + \sum_{m=1}^d \sum_{n=1}^d \Big(
%     \big (\boldsymbol{\rm e}_{k}^{(m)} \boldsymbol{\rm e}_{k}^{(n)} \big)
%     \big( \sum_{u \in \mathcal{B}} \boldsymbol{\rm e}_{u,k}^{(m)} \boldsymbol{\rm e}_{u,k}^{(n)} \big)
%     \big( \sum_{i \in \mathcal{I}} \boldsymbol{\rm e}_{i,k}^{(m)} \boldsymbol{\rm e}_{i,k}^{(n)} \big)
%     \Big),
% \end{split}
% \end{align}
\noindent where ${\hat x}_{u,i}^k$ denotes the estimated probability of user $u$ interacting with item $i$ under behavior $k$; $\mathcal{I}_{u}^{k+}$ represents the interacted items of user $u$ under behavior $k$ and $c_i^{k+}, c_i^{k-}$ are two hyperparameters.
Then the loss of the  main supervised recommendation task is the weighted sum of recommendation loss under each behavior with $\lambda_k$ as the coefficient:
\vspace{-0.3cm}
\begin{spacing}{1}
\begin{equation}
    \begin{aligned}
    \mathcal{L}_{rec} = \sum_{k=1}^K \lambda_k \mathcal{L}_{rec}^k.
    \end{aligned}
\end{equation}
\end{spacing}

% \begin{align}
%     \mathcal{L}_{rec} = \sum_{k=1}^K \lambda_k \mathcal{L}_{rec}^k.
% \end{align}

In our framework, we aim to boost the recommendation performance through customized self-supervised learning tasks, so the ultimate learning objective is the combination of main supervised task loss and SSL task loss, which is defined as:
\vspace{-0.3cm}
\begin{spacing}{1}
\begin{equation}
    \begin{aligned}
    \begin{split}
\label{loss}
    \mathcal{L} & \Rightarrow \mathcal{L}_{rec} + \mathcal{L}_{ssl1} + \mathcal{L}_{ssl2}  \\
     & \Rightarrow \mathcal{L}_{rec} + \mu_1\mathcal{L}_{ssl1}^{K,1} + ... + \mu_{K-1}\mathcal{L}_{ssl1}^{K,K-1} + \gamma\mathcal{L}_{ssl2},
\end{split}
    \end{aligned}
\end{equation}
\end{spacing}

% \begin{align}
% \begin{split}
% \label{loss}
%     \mathcal{L} & \Rightarrow \mathcal{L}_{rec} + \mathcal{L}_{ssl1} + \mathcal{L}_{ssl2}  \\
%      & \Rightarrow \mathcal{L}_{rec} + \mu_1\mathcal{L}_{ssl1}^{K,1} + ... + \mu_{K-1}\mathcal{L}_{ssl1}^{K,K-1} + \gamma\mathcal{L}_{ssl2},
% \end{split}
% \end{align}
\noindent where $\mu_k(k = 1, 2, ..., K-1)$ and $\gamma$ are hyperparameters to control the strength of the corresponding SSL task.

For one batch of the training data, the computational cost of $\mathcal{L}_{rec}$ is $O\big((|\mathcal{B}| + |\mathcal{I}|)Kd^2 + |\mathcal{E}|d\big)$, where $|\mathcal{E}|$ is the total number of positive interactions. Besides, all of the SSL task losses share the same time complexity, which equals $O\big( |\mathcal{B}|d(2+|\mathcal{V}|)\big)$.
% Therefore, the overall complexity of $\mathcal{L}$ is $O\left((|\mathcal{B}||\mathcal{V}| + |\mathcal{I}|d)Kd + |\mathcal{E}|d\right)$.
Therefore, the overall complexity of $\mathcal{L}$ is $O\left(|\mathcal{B}||\mathcal{V}|Kd + |\mathcal{E}|d\right)$, which is comparable to SOTA multi-behavior recommendation models, (\eg, CML, GHCF).
\vspace{-0.25cm}
\subsubsection{Hybrid Manipulation on Gradients}
\label{sec:hyb}
\begin{figure}[ht]
    \setlength{\abovecaptionskip}{-0.5em}
    \setlength{\belowcaptionskip}{-1.5em}
  \centering
    \subfigure[Proportion of conflicting gradients]{
        \begin{minipage}[]{0.45\linewidth}
        \includegraphics[width=1.1\linewidth]{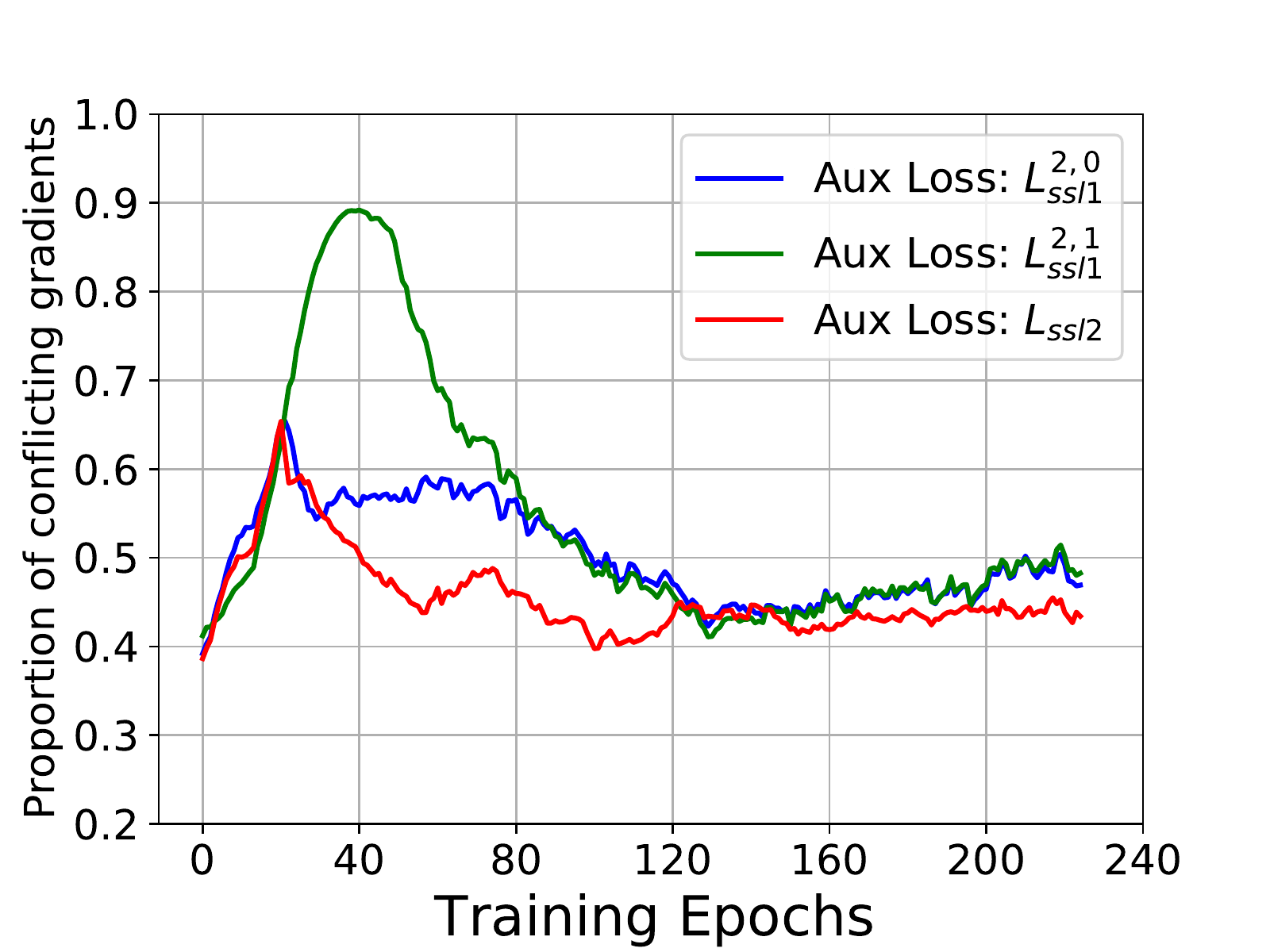}
        % \centerline{$ssl1$}
        \end{minipage}
    }
    \subfigure[Gradient Magnitude of MLP parameters]{
        \begin{minipage}[]{0.45\linewidth}
        \includegraphics[width=1.1\linewidth]{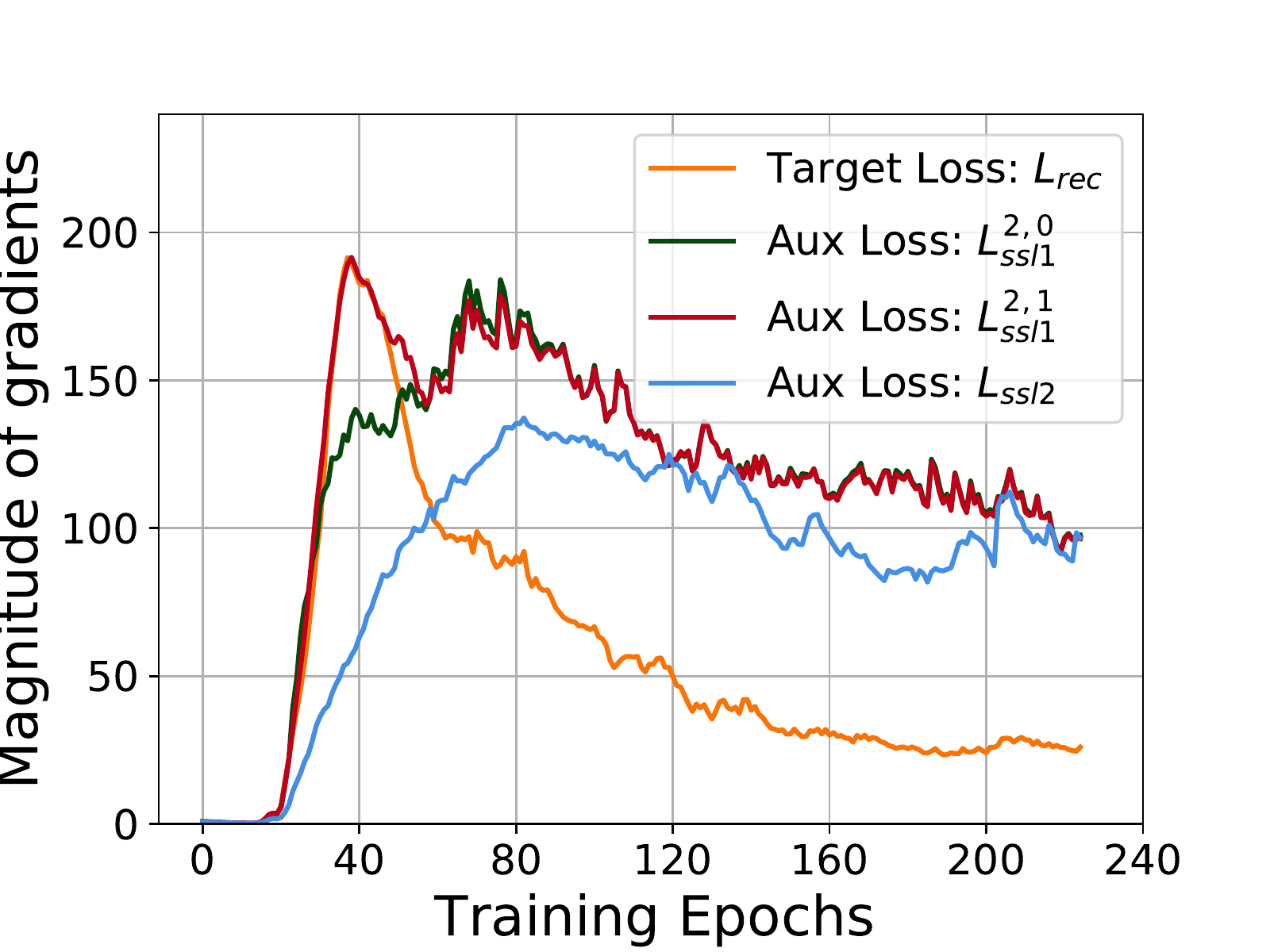}
        % \includesvg[width=\linewidth]{grad_mag.svg}
        % \centerline{$ssl1$}
        \end{minipage}
    }
  \caption{The illustration of large differences of gradient direction and magnitude between auxiliary tasks and target task.}
  \label{gradient}
\end{figure}

% hybrid manipulation on gradient magnitude and direction
Similar to Equation (\ref{loss}) where SSL tasks are viewed as auxiliary tasks to improve recommendation, existing SSL recommendation models ~\cite{liu2021contrastive, yao2021self, wu2021self, wei2022contrastive} jointly optimize the main supervised recommendation task along with the SSL task. However, they suffer from two limitations. On one hand, they neglect the potential for a significant optimization imbalance across tasks, which could deteriorate the performance of the target task. This problem becomes particularly prominent when utilizing SSL tasks to serve as auxiliary tasks because SSL pretext tasks barely coincide with the target task perfectly under a manually-designed manner. Take our MBSSL model as an example, Figure \ref{gradient}(a) and \ref{gradient}(b) highlight two examples from Beibei, which respectively demonstrate the imbalance phenomenon of gradient direction and gradient magnitude between SSL tasks and the target task.

% \todo{The figure reveals the fact that the gradient direction and magnitude of auxiliary tasks differ greatly from those of target task.}
% \todo{The figure reveals the fact that auxiliary gradients differ greatly from the target gradient in both direction and magnitude.}
On the other hand, tuning the weights of multiple auxiliary task losses (\ie, $\mu_k$, $\gamma$ in Equation (\ref{loss})) is time-consuming, and fixed weights do not apply to all the batches throughout the dynamic training process. Towards this end, we develop an adaptive optimization method for SSL recommendation models with hybrid manipulation on both the magnitude and direction of gradients.

In MBSSL, each SSL task loss (\ie, $\mathcal{L}_{ssl1}^{K,1}, ... \mathcal{L}_{ssl1}^{K,K-1}, \mathcal{L}_{ssl2}$) can be viewed as an auxiliary task loss denoted as $\mathcal{L}_{aux, i}$ while the main supervised task loss can be rewritten as the target task loss $\mathcal{L}_{tar}$. Let $\theta$ denote the set of bottom shared parameters; $t$ denotes the $t$-th training iteration within one epoch;
% $\boldsymbol{\rm G}_{tar}^t=\nabla_\theta \mathcal{L}_{tar}^t $
$\boldsymbol{\rm G}_{tar}^t,  \boldsymbol{\rm G}_{aux,i}^t$ respectively denote the gradient of the target task and auxiliary task with respect to $\theta$, \ie, $\boldsymbol{\rm G}_{tar}^t = \nabla_\theta \mathcal{L}_{tar}^t$ and $\boldsymbol{\rm G}_{aux, i}^t = \nabla_\theta \mathcal{L}_{aux,i}^t$. Hence, our goal is to  balance $\boldsymbol{\rm G}_{aux,i}^t$ and $\boldsymbol{\rm G}_{tar}^t$ during each iteration via modifying the direction and magnitude of $\boldsymbol{\rm G}_{aux,i}^t$ adaptively.

Intuitively, gradient with larger magnitude will dominate the
optimization trend, so in the case where $||\boldsymbol{\rm G}_{aux, i}^t|| \geq  ||\boldsymbol{\rm G}_{tar}^t||$, the
optimizer prefers to approach the $i$-th auxiliary task rather than
the target task which leads to the performance degradation. As a
solution, we dedicate to alter the direction and magnitude of auxiliary
gradient $\boldsymbol{\rm G}_{aux, i}^t$ which embraces larger magnitude than the target task $\boldsymbol{\rm G}_{tar}^t$, so as to guide the optimization towards the target task. In terms of auxiliary gradients with smaller magnitude as well as conflicting directions, we keep them unaltered for preventing overfitting. To be more specific, we first modify the gradient direction via projecting the auxiliary gradient to the normal plane of the target gradient if they conflict with each other, \ie, their cosine similarity is negative. The projection strategy is formulated as:
\vspace{-0.3cm}
\begin{spacing}{1}
\begin{equation}
    \begin{aligned}
    \boldsymbol{\rm G}_{aux, i} ^t =  \boldsymbol{\rm G}_{aux, i} ^t - \frac{\boldsymbol{\rm G}_{aux, i} ^t \cdot \boldsymbol{\rm G}_{tar} ^t}{||\boldsymbol{\rm G}_{tar} ^t||^2} \boldsymbol{\rm G}_{tar} ^t.
    \end{aligned}
\end{equation}
\end{spacing}

Although the amount of destructive gradient interference has been reduced via direction modification, large gradient magnitudes of auxiliary tasks still hinder the optimization towards the target task. Therefore, we further balance the gradient magnitude based on ~\cite{he2022metabalance, malkiel2020mtadam} with a relax factor $r$ to control the magnitude proximity between $\boldsymbol{\rm G}_{aux, i}$ and $\boldsymbol{\rm G}_{tar}$:

% \vspace{-1cm}
\begin{spacing}{1}
\begin{equation}
\begin{aligned}
     \boldsymbol{\rm G}_{aux, i} ^t =  r *  \frac{||\boldsymbol{\rm G}_{tar} ^t||}{||\boldsymbol{\rm G}_{aux, i} ^t ||} \boldsymbol{\rm G}_{aux, i} ^t + (1-r) * \boldsymbol{\rm G}_{aux, i} ^t.
\end{aligned}
\end{equation}
\end{spacing}

Through such a hybrid manipulation altering both the gradient direction and magnitude of auxiliary tasks, the learning process could readily optimize towards the target task, thus boosting the recommendation performance of target behavior. The complete update procedure of hybrid manipulation is shown in Algorithm \ref{alg:hmg}.

\begin{algorithm}[t]
\footnotesize
\caption{The Hybrid Manipulation on Gradients.}

\label{alg:hmg}
\LinesNumbered
\KwIn{
Initial shared parameters $\theta_0$; Relax factor $r$; Learning rate $\eta$;  Auxiliary task loss $\mathcal{L}_{aux,i} \in  \{\mathcal{L}_{ssl1}^{K,1}, ..., \mathcal{L}_{ssl1}^{K,K-1}, \mathcal{L}_{ssl2}\}$;
Target task loss $\mathcal{L}_{tar}=\mathcal{L}_{rec}$;
 }

\KwOut{Updated shared parameters $\theta_T$}
\For{$t = 1$ to $T$ }{
    $\boldsymbol{\rm G}_{tar}^t = \nabla_\theta \mathcal{L}_{tar}^t$ \\
    \For{$i = 1$ to $K$}{
        $\boldsymbol{\rm G}_{aux, i}^t = \nabla_\theta \mathcal{L}_{aux,i}^t$ \\
        \If{$||\boldsymbol{\rm G}_{aux,i}^t|| > ||\boldsymbol{\rm G}_{tar}^t||$}{
            $\boldsymbol{\rm G}_{aux, i} ^t \leftarrow  \boldsymbol{\rm G}_{aux, i} ^t - \frac{\boldsymbol{\rm G}_{aux, i} ^t \cdot \boldsymbol{\rm G}_{tar} ^t}{||\boldsymbol{\rm G}_{tar} ^t||^2} \boldsymbol{\rm G}_{tar} ^t$

            $\boldsymbol{\rm G}_{aux, i} ^t \leftarrow  r *  \frac{||\boldsymbol{\rm G}_{tar} ^t||}{||\boldsymbol{\rm G}_{aux, i} ^t ||} \boldsymbol{\rm G}_{aux, i} ^t + (1-r) * \boldsymbol{\rm G}_{aux, i} ^t$
        }
    }
    $\boldsymbol{\rm G}^t \leftarrow \boldsymbol{\rm G}_{tar}^t + \boldsymbol{\rm G}_{aux, 1}^t+...+\boldsymbol{\rm G}_{aux, K}^t$\\
    $\theta_{t} \leftarrow \theta_{t-1} - \eta * \boldsymbol{\rm G}^t$
}
\Return{$\theta^T$}

\end{algorithm}

% \setlength{\textfloatsep}{0pt}
% \subsection{\review{Time Complexity}}
% \review{TODO}

% \enlargethispage{1em}
% \vspace{-1cm}
\subsection{Model Analysis}
\label{sec:method_comp}
In this section, we compare the proposed SSL paradigm with that of existing representative work on multi-behavior recommendation.

\noindent \textbf{Comparison with CML.} Likewise, the cross-behavior SSL  in CML is performed between each auxiliary and target behavior pair to capture cross-type behavior dependency. Specifically, the SSL paradigm follows the conventional rule, \ie, the views of any different users will be regarded as negative pairs. However, we can conclude that users may share similar preferences based on rich semantics and huge data volume of behaviors, which means the common practice may lead to many false negative pairs. Therefore, in our inter-behavior SSL, we selectively construct negative pairs based on the calculated structural node similarities to facilitate the knowledge transfer between auxiliary and target behaviors.

\noindent \textbf{Comparison with S-MBRec.} The star-style SSL in S-MBRec constructs additional positive samples by finding similar users based on the data under target behavior. However, the data is so sparse that the calculated node similarities are not reliable. What's worse, the negative transfer will be further amplified under the current SSL paradigm which encourages the alignment between unreliable positive samples. Accordingly, we aim to make full use of data under all the behaviors to select potential similar users with a high confidence level, and refuse to augment positive samples accounting for the robustness against interaction noises \cite{chen2021incremental, huynh2022boosting}.

All of the existing work solely rely on the inter-behavior SSL to handle the data sparsity issue, which is not enough though. And the inter-behavior SSL is likely to introduce noises from auxiliary behaviors. As a solution, we conduct intra-behavior SSL within the target behavior itself with an aim to counteract the auxiliary noises via amplifying the impact of the target behavior.

% \enlargethispage{2em}
% \vspace{-0.5cm}
\section{Experiments}
\label{sec:exp}
In this section, we conduct extensive experiments on five real-world datasets to evaluate our proposed model. In particular, we aim to figure out the following research questions: \\
$\bullet$ \textbf{RQ1}: How effective is MBSSL in multi-behavior recommendation scenarios compared to existing methods? \\
$\bullet$ \textbf{RQ2}: How do the sub-modules of MBSSL affect the recommendation performance? \\
$\bullet$ \textbf{RQ3}: How robust is MBSSL to data sparsity under the target behavior and to noisy interactions from auxiliary behaviors?\\
$\bullet$ \textbf{RQ4}: Which strategy of combining the manipulation on gradient direction and magnitude is the best?\\
$\bullet$ \textbf{RQ5}: What is the influence of different hyperparameter settings on recommendation performance?\\

% \begin{itemize}[leftmargin=*]
%     \item \textbf{RQ1}: How effective is MBSSL in multi-behavior recommendation scenarios compared to existing works?
%     \item \textbf{RQ2}: How do the sub-modules of MBSSL affect the recommendation performance?
%     \item \textbf{RQ3}: How robust is MBSSL to data sparsity under the target behavior and to noisy interactions from auxiliary behaviors?
%     \item \textbf{RQ4}: Which strategy of combining the manipulation on gradient direction and magnitude is the best?
%     \item \textbf{RQ5}: What is the influence of different hyperparameter settings on recommendation performance?

% \end{itemize}
\vspace{-0.5cm}
\subsection{Experimental Settings}

\begin{table}
\footnotesize
\centering
\caption{The statistics of datasets.}
\setlength{\abovecaptionskip}{10em}
\setlength{\belowcaptionskip}{-10em}
\begin{tabular}{ccccc}
\hline
Dataset & \#User & \#Item & \makebox[0.001\textwidth][c]{\#Interactions} & \makebox[0.001\textwidth][c]{Interaction Behavior Type} \\
\hline
Beibei & 21,716& 7,977 & 3,338,068 & \{Page View, Cart, Purchase\}  \\
Taobao & 48,749 & 39,493 & 1,952,931 & \{Page View, Cart, Purchase\}  \\
Tmall & 31,882 & 31,232 & 1,451,219 & \{Page View, Favorite, Cart, Purchase\}  \\
IJCAI-Contest & 17,435 & 35,920 & 799,368 &\{Page View, Favorite, Cart, Purchase\}  \\
Videos & 29,197 & 23,251 & 2,002,201 & \{Click, Like, Comment, Download \}  \\
% Kuaishou & 138,812 & 1,779,639 & 2 & 4 & Yes & Yes \\
\hline
\end{tabular}
\label{tab:datasets}
% \vspace{-0.5cm}
\end{table}
\subsubsection{Datasets.}
% To evaluate the performance of MBSSL, we experiment on four public datasets \ie, Beibei, Taobao, Tmall, IJCAI-Contest, and one internal dataset, Videos. Note that We summarize the statistics of all the datasets in Table \ref{tab:datasets}.

To evaluate the performance of MBSSL, we experiment on four public datasets \ie, Beibei, Taobao, Tmall, IJCAI-Contest which are publicly available split datasets and one internal dataset, Videos. Table \ref{tab:datasets} summarizes the statistics of all the datasets.

% \footnote{https://www.beibei.com}
% \footnote{https://tianchi.aliyun.com/dataset/dataDetail?dataId=649}
\begin{itemize}[leftmargin=*]
    \item \textbf{Beibei}. This is the dataset collected from Beibei, the largest infant product e-commerce platform in China. It involves three types of interaction behavior, \ie, \textit{page view, cart, purchase}, among which \textit{purchase} is the target behavior.
    \item\textbf{Taobao}. This is an open dataset obtained from the largest e-commerce site Taobao, which contains the same interaction type with Beibei. We directly use the processed dataset in GHCF\cite{chen2021graph}.

    \item\textbf{Tmall}. This is another processed dataset on Taobao provided by CML ~\cite{wei2022contrastive}, which contains one additional behavior \textit{favorite}.

    \item \textbf{IJCAI-Contest}. This dataset is provided in IJCAI15 Challenge. It is collected from a business-to-customer retail system, which shares the same behavior types with the Tmall data.
    \item \textbf{Videos}. This is a dataset collected from an online short video platform. There are totally four types of interaction behavior between users and videos, \ie, \textit{click, like, comment, download}. In this dataset, we regard \textit{download} as the target behavior.
\end{itemize}

\begin{table*}
\footnotesize
\caption{The performance comparison on five datasets. Note that baselines with the "all" suffix use data from all the behaviors to build the single-behavior model. The best results are illustrated in bold and the number underlined is the runner-up. And the number with a star ($*$) indicates the result is statistically evaluated with $p < 0.05$ under t-test compared to other baselines.}
\centering
\centering
\begin{tabular}{c|cccc|cccc|cccc}
\hline
& \multicolumn{4}{c|}{Beibei} & \multicolumn{4}{c}{Taobao} & \multicolumn{4}{c}{Tmall}  \\
& R@10 & R@50  & N@10 & N@50  & R@10 & R@50  & N@10 & N@50 & R@10 & R@50  & N@10 & N@50 \\
\hline
NGCF ~\cite{wang2019neural} & 0.0268 & 0.1008 & 0.0124 &  0.0281 &  0.0135 & 0.0330 & 0.0073 & 0.0115 & 0.0155	& 0.0399	& 0.0078 & 0.0131 \\
LightGCN ~\cite{he2020lightgcn}  &  0.0383 &	0.1366	&0.0190 & 0.0399 & 0.0163 & 0.0477 & 0.0085 & 0.0152	& 0.0172&	0.0479&	0.0078	&0.0144 \\
NGCF\_all ~\cite{wang2019neural} & 0.0286 & 0.1084 & 0.0139 &  0.0308 &  0.0144 & 0.0391 & 0.0075&  0.0127& 0.0341 &	0.0752 &	0.0183	& 0.0272\\
LightGCN\_all ~\cite{he2020lightgcn}  &  0.0552 & 0.1626 & 0.0284&  0.0514 & 0.0405 & 0.0991 & 0.0223 &  0.0350 & 0.0328	& 0.0821	& 0.0174	& 0.0279\\
\hline
SGL ~\cite{wu2021self} & 0.0418 & 0.1383 & 0.0201 & 0.0405 &  0.0431&0.0775&0.0257&0.0332 & 0.0340&	0.0700	&0.0185	& 0.0264 \\
SimGCL ~\cite{yu2022graph} & 0.0479 & 0.1548 & 0.0231 & 0.0460 & 0.0395 & 0.0772&  0.0226 & 0.0308  & 0.0347 & 0.0775 &  \underline{0.0193} & 0.0286 \\
SGL\_all ~\cite{wu2021self} & 0.0597 & 0.1699 & 0.0310 &  0.0545 &  0.0506 & 0.1214 & 0.0278 & 0.0430 & 0.0281 & 0.0779 & 0.0148 & 0.0255 \\
SimGCL\_all ~\cite{yu2022graph}  & 0.0603 & 0.1697 & 0.0311 & 0.0545 & 0.0502 & 0.1242 & 0.0274 & 0.0433 & 0.0327 & 0.0903 & 0.0166 & 0.0290 \\
\hline
MATN ~\cite{xia2020multiplex}  &   0.0466 & 0.1224 & 0.0261 &  0.0422 &  0.0259 & 0.0760 & 0.0132 &  0.0239 & 0.0154 & 0.0558 & 0.0075 & 0.0161 \\
MBGMN ~\cite{xia2021graph}& 0.0427 & 0.1439 & 0.0227&  0.0441 &  0.0485 & 0.1314 & 0.0250 &  0.0428 & 0.0263	& 0.0741 &	0.0136 &0.0239 \\
CML ~\cite{wei2022contrastive}& 0.0588 & 0.1953 & 0.0292 &  0.0584 &  0.0299& 0.0914&0.0150&0.0282	&0.0061	&0.0242	& 0.0027	& 0.0065\\
EHCF ~\cite{chen2020efficient}& 0.1523 & 0.3316 & 0.0817 &  0.1213 &  0.0717	&0.1618	&0.0403&	0.0594&	0.0234	&0.0642&	0.0116	&0.0204\\
S-MBRec ~\cite{gu2022self} & 0.1697 & 0.3708 & 0.0872 & 0.1313 & \underline{0.0814} & 0.1878 & \underline{0.0446} & 0.0677 & 0.0261 & 0.0771 & 0.0125 & 0.0235 \\
GHCF ~\cite{chen2021graph}& \underline{0.1922} & \underline{0.3794} & \underline{0.1012} &  \underline{0.1426} &  0.0807	& \underline{0.1892}	&0.0442&	\underline{0.0678}	&\underline{0.0353}	&\underline{0.0954}	&0.0175&	\underline{0.0303}
 \\

\hline
MBSSL &  \textbf{0.2229}$^{*}$ & \textbf{0.3806}$^{*}$ & \textbf{0.1277}$^{*}$ & \textbf{0.1626}$^{*}$ & \textbf{0.1027}$^{*}$ & \textbf{0.2120}$^{*}$ & \textbf{0.0576}$^{*}$ &  \textbf{0.0813}$^{*}$  & \textbf{0.0400}$^{*}$ & \textbf{0.1101}$^{*}$ & \textbf{0.0201}$^{*}$ &  \textbf{0.0352}$^{*}$ \\
\hline
\end{tabular}

~

\begin{tabular}{c|cccc|cccc}
\hline
& \multicolumn{4}{c|}{IJCAI-Contest} & \multicolumn{4}{c}{Videos} \\
& R@10 & R@50  & N@10 & N@50  & R@10 & R@50  & N@10 & N@50 \\
\hline
NGCF ~\cite{wang2019neural}& 0.0041 & 0.0115 & 0.0026&  0.0042 &  0.0178  & 0.0666 & 0.0084 &  0.0187 \\
LightGCN ~\cite{he2020lightgcn}& 0.0143 & 0.0323 & 0.0076 &  0.0113 & 0.0307 & 0.0803 & 0.0158 &  0.0265\\
NGCF\_all ~\cite{wang2019neural}& 0.0102 & 0.0254 & 0.0056&  0.0089 &  0.0808  & 0.1995 & 0.0449 &  0.0704\\
LightGCN\_all ~\cite{wang2019neural}& 0.0274 & 0.0669 & 0.0156 &  0.0244 &  0.0887  & 0.2060 & 0.0491 &  0.0745\\
\hline
SGL  ~\cite{wu2021self}&  0.0157 & 0.0323 & 0.0086 & 0.0120
 & 0.0337 & 0.0832 & 0.0179 & 0.0420 \\
SimGCL ~\cite{yu2022graph}& 0.0156 & 0.0360 & 0.0083 & 0.0127 & 0.0371 & 0.0964 & 0.0199 & 0.0326 \\
SGL\_all ~\cite{wu2021self} & 0.0269 & 0.0644 & 0.0151 &  0.0232 &  0.1002 &  0.2179 & 0.0575 & 0.0830 \\
SimGCL\_all ~\cite{yu2022graph} & 0.0283 & 0.0710 & 0.0163 & 0.0254 & 0.1053 & 0.2282 & 0.0595 & 0.0860 \\
\hline
MATN ~\cite{xia2020multiplex}& 0.0113&0.0347&0.0053&	0.0102
 &  0.0053  &  0.0373 &  0.0023 &  0.0089 \\
MBGMN ~\cite{xia2021graph}& 0.0305 & 0.0647 & 0.0172 & 0.0246
 &  0.0365 & 0.0786 & 0.0220 &  0.0311 \\
CML ~\cite{wei2022contrastive}& 0.0249 & 0.0483 & 0.0142 & 0.0192 &  0.0011 & 0.0056  & 0.0005&  0.0020 \\
EHCF ~\cite{chen2020efficient} & 0.0264 & 0.0552 & 0.0152 & 0.0215 &  0.1036 & 0.2947 & 0.0530&  0.0941 \\
S-MBRec ~\cite{gu2022self} & 0.0292 & 0.0805 & 0.0160 & 0.0268 & 0.0911 & 0.1944 & 0.0461 & 0.0690 \\
GHCF  ~\cite{chen2021graph}&  \underline{0.0305} &\underline{ 0.0830} & \underline{0.0168} &  \underline{0.0281} & \underline{0.1687} & \underline{0.3512} & \underline{0.0808} &  \underline{0.1220}\\

\hline
MBSSL & \textbf{0.0390}$^{*}$ & \textbf{0.0952}$^{*}$ & \textbf{0.0201}$^{*}$ &  \textbf{0.0321}$^{*}$ &  \textbf{0.1835}$^{*}$ & \textbf{0.3768}$^{*}$ & \textbf{0.0869}$^{*}$ &  \textbf{0.1306}$^{*}$ \\
\hline
\end{tabular}

\label{tab:exp-link}
\end{table*}

\subsubsection{Baselines.}
We compare our method with the following state-of-the-art methods from three types: Single-Behavior, Self-supervised Learning and Multi-Behavior recommendation models. For Single-Behavior and Self-supervised Learning methods, we normally utilize target behavior data in the same way as CML ~\cite{wei2022contrastive} or GHCF  ~\cite{chen2021graph} to build the model. However, in order to eliminate the performance gap resulting from different volumes of training data, we additionally conduct experiments on these models by treating all the behaviors as the same type, which leads to stronger baselines.

\noindent \textbf{Single-Behavior Recommendation Models.}
\begin{itemize}[leftmargin=*]
    \item \textbf{NGCF} ~\cite{wang2019neural}: It is a neural collaborative filtering method utilizing GNNs, with an aim to capture high-order connections.
    \item \textbf{LightGCN} ~\cite{he2020lightgcn}: It simplifies the GCN structure to improve training efficiency and generalization ability for recommendation.
\end{itemize}

\noindent \textbf{Self-supervised Learning Recommendation Models.}
\begin{itemize}[leftmargin=*]
    % \item \textbf{NMTR}

    % \item \textbf{DNN+SSL} ~\cite{yao2021self}: It is the first state-of-the-art SSL solution for recommendation. It uses the classical two-tower architecture where DNN is the feature encoder together with an SSL module conducting mask or dropout on item features.
    \item \textbf{SGL} ~\cite{wu2021self}: This method explores SSL on graph structure and accordingly devises three unified augmentation operators including node dropout, edge dropout and random walk.
    \item \textbf{SimGCL} ~\cite{yu2022graph}: It is a simple yet effective graph-augmentation-free contrastive learning method that can regulate the uniformity in a smooth way.
\end{itemize}

\noindent \textbf{Multi-Behavior Recommendation Models.}
\begin{itemize}[leftmargin=*]
    % \item \textbf{NMTR}
    \item \textbf{MATN}~\cite{xia2020multiplex}: It proposes an attention-based transformer encoder to help preserve cross-type behavior collaborative signals and type-specific behavior contextual information.
    \item \textbf{MBGMN} ~\cite{xia2021graph}: It enhances multi-behavior modeling with Graph Meta Network which incorporates the meta learning paradigm.
    \item \textbf{CML}~\cite{wei2022contrastive}: It designs a contrastive learning framework for multi-behavior recommendation and further utilizes meta learning to learn the customized weights for each user.
    \item \textbf{EHCF} ~\cite{chen2020efficient}: It conducts knowledge transfer among behaviors and proposes a novel non-sampling objective for multi-behavior recommendation.
    \item \textbf{S-MBRec} ~\cite{gu2022self}: It is another SSL-based model which considers the discrepancies
and commonalities of multiple behaviors.
    \item \textbf{GHCF} ~\cite{chen2021graph}: It is an improvement over EHCF which relies on the GNNs to model the complex high-hop user-item correlations.
\end{itemize}

\vspace{-0.2cm}
\subsubsection{Evaluation Methodology.} For a fair comparison with various models on recommendation, we adopt the widely-used leave-one-out evaluation and two ranking metrics, \textit{Recall@K} and \textit{NDCG@K}. Note that we utilize the same evaluator as EHCF ~\cite{chen2021graph} or GHCF ~\cite{chen2020efficient}, \ie, we rank all the items except positive ones for each user, which is more persuasive than randomly sampling a subset of non-interactive items for each user (\eg, in MBGMN ~\cite{xia2021graph} or CML ~\cite{wei2022contrastive}, pairing each positive item instance with 99 randomly sampled items).
% \begin{itemize}[leftmargin=*]
%     \item \textbf{Recall@K}: It measures the ratio of test items which are included in the Top-K list.
%     \item \textbf{NDCG@K}: It is a position-sensitive metric which assigns higher scores to hits at higher positions in the Top-K list.
% \end{itemize}
% \enlargethispage{1em}

\vspace{-0.2cm}
\subsubsection{Parameter Settings.}
We implement our MBSSL model with Pytorch and the model is optimized using the Adam optimizer with learning rate 0.001 during the training phase. We set $\alpha$ to 0.5 for swing algorithm. The batch size ranges from \{256, 512, 1024\}. By default, the size of the latent dimension is set as 64 and the number of propagation layers is 4. In addition, the negative weight of the non-sampling loss is chosen from \{0.1, 0.01\} for all the datasets. For preventing overfitting, we set the embedding dropout ratio as 0.3 and the edge dropout ratio $\rho$ as 0.5.

\vspace{-0.2cm}
\subsection{Performance Comparison (RQ1)}
In the evaluation, we first perform experiments to make recommendations on five datasets where we set the recommendation length $K$ as 10, 50. From Table \ref{tab:exp-link}, we summarize the following observations:

\begin{itemize}[leftmargin=*]
    \item Our model MBSSL generally outperforms all the baselines on all datasets, with the improvements over the best baseline ranging from 18.64\% to 19.71\% in terms of Recall@10 and NDCG@10. The significant improvements can be attributed to two main reasons: i) Through inter-behavior and intra-behavior SSL, our model effectively addresses the data sparsity under the target behavior and noisy interactions from auxiliary behaviors, which are two key problems of multi-behavior recommendation. ii) The proposed hybrid manipulation method on gradients empowers the capability of balancing the optimization between the SSL task and the main supervised recommendation task, which further preserves the recommendation performance.

    \item Interestingly, we find that some multi-behavior models (\eg, MATN, MBGMN) may perform worse than some single-behavior models which treat all the behaviors as the same type. (\eg, LightGCN). The poorer performance of MBGMN and MATN may suggest their disadvantages of differentiating behavior types. Whereas, methods like EHCF, GHCF and MBSSL are superior to single-behavior models in general, which highlights the significance of behavior modeling.

    \item SSL based models (\ie, SGL and SimGCL) yield better performance than single-behavior models, which indicates that exploiting SSL on graph-structure could empower the generalization ability of GNN-based recommender models. However, the optimization imbalance of the SSL task and the recommendation task remains unexplored and this may lead to a performance decline.
\end{itemize}

% \enlargethispage{2em}
\vspace{-0.3cm}
\subsection{Ablation Study (RQ2)}
In this subsection, we evaluate the rationality of each designed module in our model, with four variants as follows:
\begin{itemize}[leftmargin=*]
    \item \textbf{\textit{w/o} \boldmath{$CDM$}}: The cross-behavior dependency modeling is removed in the behavior-aware graph neural network.
    \item \textbf{\textit{w/o} \boldmath{$SSL_{inter}$}}: We do not perform inter-behavior SSL in the multi-behavior self-supervised learning part.
    \item \textbf{\textit{w/o} \boldmath{$SSL_{intra}$}}: We remove intra-behavior SSL in the multi-behavior self-supervised learning part.
    \item \textbf{\textit{w/o} \boldmath{$HMG$}}: We disable the hybrid manipulation on gradients, \ie, we do not tackle the optimization imbalance. Instead, we assign equal weights for each auxiliary loss.
\end{itemize}

The ablation study results are shown in Table \ref{tab:ab}. Note that due to space limitations, we only show the results of Recall@10 and NDCG@10 on two datasets, Beibei and Taobao. For the other datasets, the observations are similar. From Table \ref{tab:ab}, we can find:
\begin{itemize}[leftmargin=*]
    \item The cross-behavior dependency modeling plays a vital role in performance, which indicates that the self-attention mechanism has a strong capability on capturing the implicit pair-wise dependencies across behaviors.

    \item Removing either part of the multi-behavior self-supervised learning will undermine the performance. Specifically, the performance gap between \textit{MBSSL} and \textit{w/o} $SSL_{inter}$ demonstrates the effectiveness of the inter-behavior SSL on narrowing the gap of skewed representations and alleviating the data sparsity of the target behavior. In addition, the intra-behavior SSL contributes to the performance further, indicating the necessity to address the noisy interactions from auxiliary behaviors.

    \item When we replace the HMG by assigning equal weights for all the auxiliary losses, the performance experiences a great decline which verifies the existence of the optimization imbalance. This also suggests that during the multi-behavior learning process, HMG adaptively rectifies the gradients to balance the auxiliary tasks, which  improves the target task’s performance significantly.
\end{itemize}

\begin{table}
% \scriptsize
% \setlength{\belowcaptionskip}{-1em}
\footnotesize
\caption{The experimental results of ablation study.}
\setlength{\abovecaptionskip}{-2em}
\setlength{\belowcaptionskip}{-1em}
\label{tab:ab}

\centering
\begin{tabular}{c|cc|cc}
% \toprule
\hline
Data & \multicolumn{2}{c|}{Beibei} & \multicolumn{2}{c}{Taobao} \\
Metric  & Recall@10 & NDCG@10 & Recall@10 & NDCG@10 \\
\hline
\hline
\textit{w/o} $CDM$ & 0.1923 & 0.1059  & 0.0936 & 0.0520 \\
\textit{w/o} $SSL_{inter}$ & 0.2025 &  0.1108 & 0.0902 & 0.0515 \\
\textit{w/o} $SSL_{intra}$ & 0.1994 & 0.1027 &  0.0943 & 0.0526   \\
\textit{w/o} $HMG$ & 0.2086 & 0.1131 & 0.0936 & 0.0530 \\
\hline
\hline
MBSSL & \textbf{0.2229}  & \textbf{0.1277} & \textbf{0.1027} & \textbf{0.0576} \\
\hline
\end{tabular}
% \vspace{-0.5cm}
\end{table}

% \enlargethispage{2em}
\vspace{-0.3cm}
\subsection{Robustness Analysis (RQ3)}
\subsubsection{Robustness to data sparsity}
In recommender systems, the recommendation towards  inactive users with few available interactions is quite challenging, so we aim to illustrate the effectiveness of our model in alleviating the data sparsity. Specifically, we split the test users into five groups based on sparsity degrees, \ie, the number of interactions under the target behavior (\ie, purchase), then we compute the average NDCG@50 on each group of users.

\begin{figure}[t]
  \centering
    \setlength{\belowcaptionskip}{-1em}
    % \subfigure[Beibei Recall@50]{
    %     \begin{minipage}[]{0.45\linewidth}
    %     \includegraphics[width=\linewidth]{sparse_bb_recall.pdf}
    %     \end{minipage}
    % }
    \subfigure[Beibei NDCG@50]{
        \begin{minipage}[]{0.45\linewidth}
        \includegraphics[width=\linewidth]{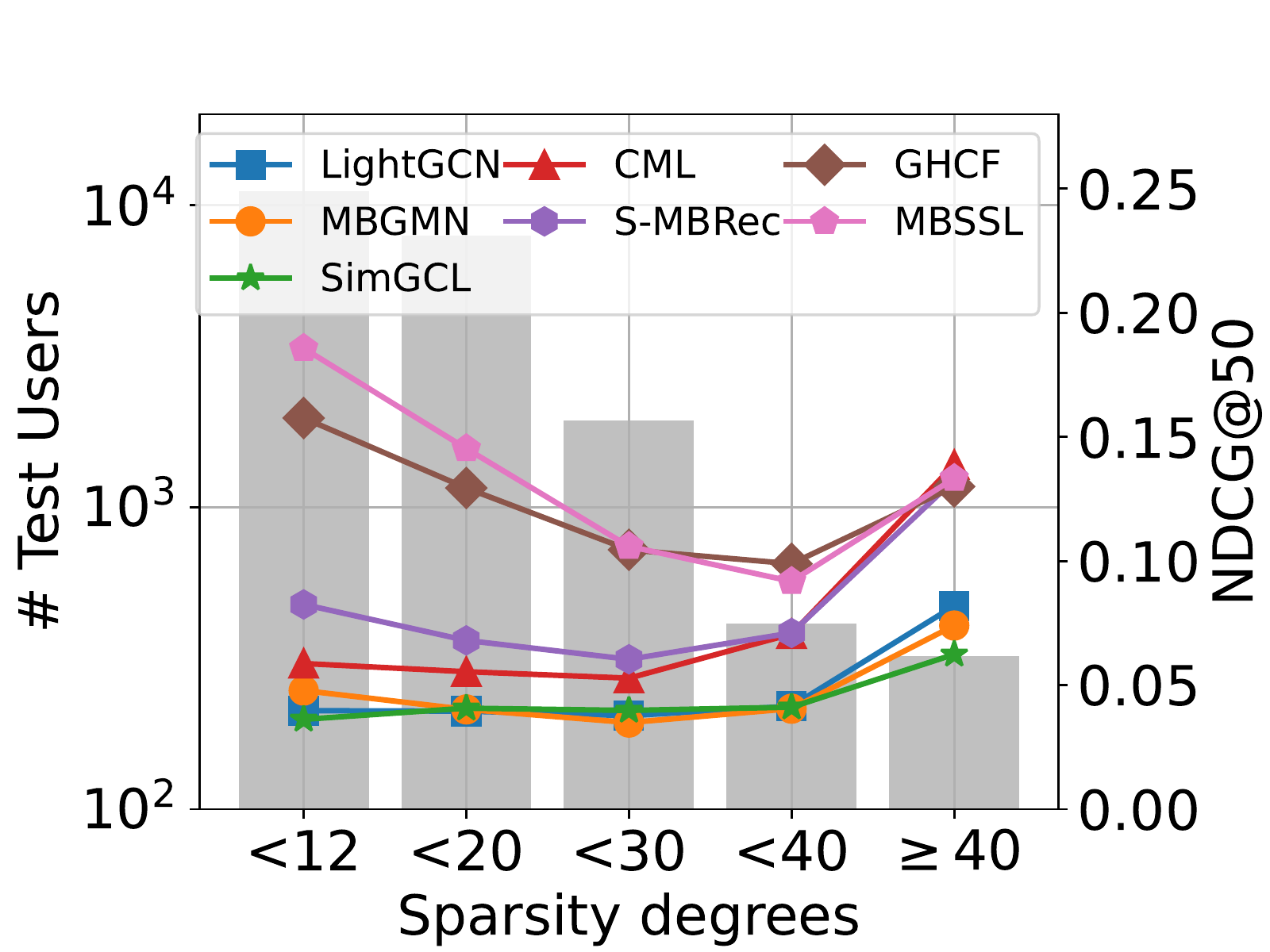}
        \end{minipage}
    }
    % \vspace{-0.4cm}
    % \subfigure[Tmall Recall@50]{
    %     \begin{minipage}[]{0.45\linewidth}
    %     \includegraphics[width=\linewidth]{sparse_tb_recall.pdf}
    %     % \centerline{$ssl1$}
    %     \end{minipage}
    % }
    \subfigure[Taobao NDCG@50]{
        \begin{minipage}[]{0.45\linewidth}
        \includegraphics[width=\linewidth]{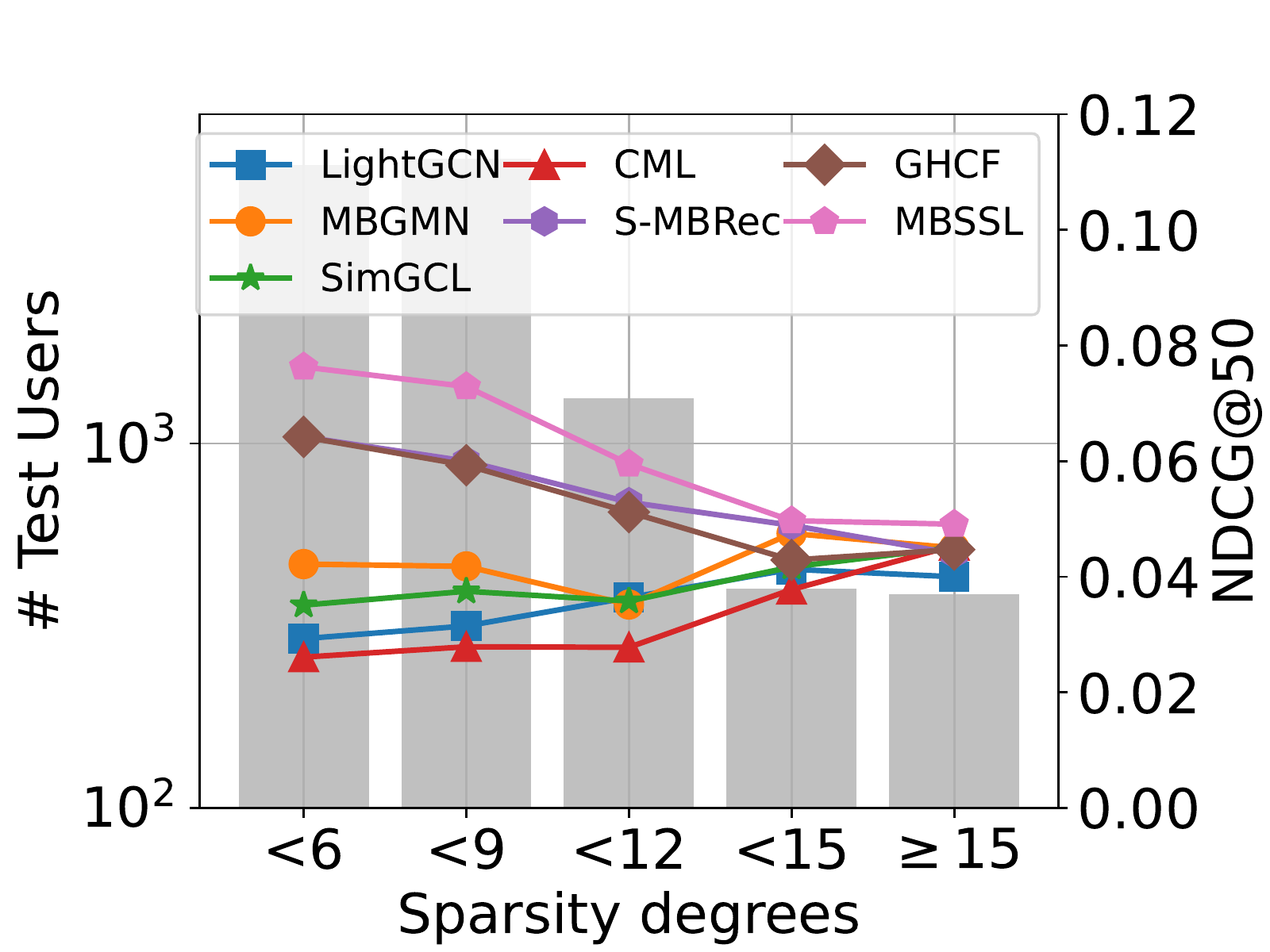}
        \end{minipage}
    }
  \caption{Performance comparison \textit{w.r.t} different data sparsity degrees on Beibei and Taobao.}
  \label{sparsity}

\end{figure}

Due to space limitations, we present the comparison results with representative baselines on two public datasets Beibei and Taobao, results on other datasets would be later attached if needed. In Figure \ref{sparsity}, the x-axis denotes the different data sparsity degrees, the left side of y-axis displays the number of test users in the corresponding group quantified by bar while the right side of y-axis is the averaged metric value quantified by line.

Based on the results, we have the following observations: i) Our model generally obtains a better performance compared to other SOTA methods on these two datasets. On Beibei dataset, despite the slight inferiority to GHCF and CML on part of active users, our model manifests a  good capability of recommendation for inactive users who occupy a considerable amount of user populations.
% ii) Multi-behavior recommendation models are generally superior to single-behaivor models on Beibei while on Tmall, some multi-behavior models like MBGMN may perform better, which uncovers the fact that they are insufficient in capturing discriminative semantics of different behaviors.
ii) The performance will encounter a slight descent when the number of interactions increases, and this may be caused by different amounts of auxiliary data. For example, on Taobao dataset, the number of auxiliary behavioral records for users who have more than 12 purchase records is much fewer than for users who have less than 9 purchase records.
% From another perspective, the results could indicate the effectiveness of auxiliary data.

\vspace{-0.2cm}
\subsubsection{Robustness to noisy interactions}
\begin{figure}[t]
  \centering
    \setlength{\belowcaptionskip}{-0.5cm}
    \subfigure[Beibei]{
        \begin{minipage}[]{0.45\linewidth}
        \includegraphics[width=\linewidth]{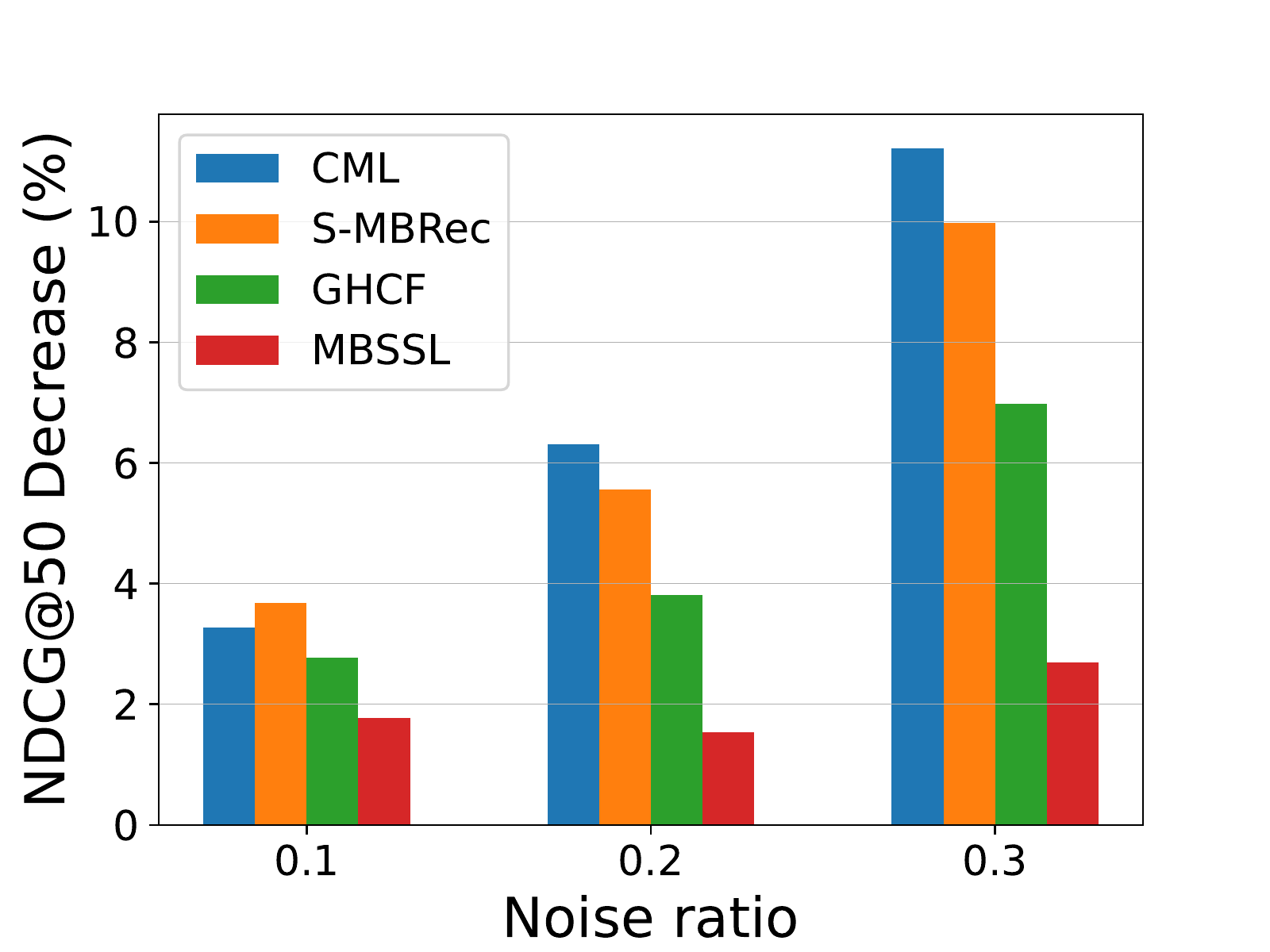}
        \end{minipage}
    }
    \subfigure[Taobao]{
        \begin{minipage}[]{0.45\linewidth}
        \includegraphics[width=\linewidth]{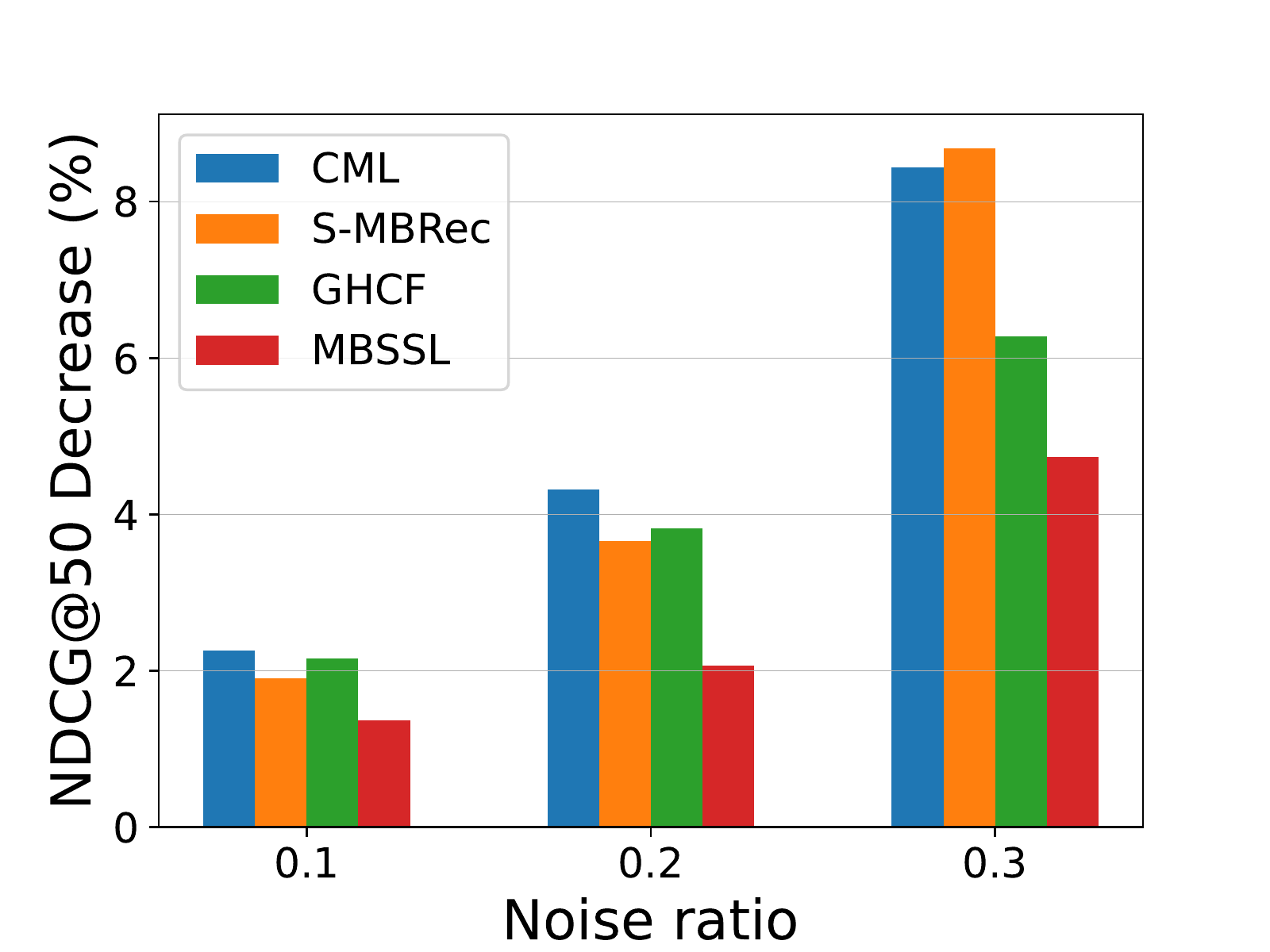}
        \end{minipage}
    }
  \caption{Performance \textit{w.r.t} noise ratio. The bar denotes the decrease percentage of the performance reported in Table \ref{tab:exp-link}.}
  \label{noise}

\end{figure}

As mentioned above, the inter-behavior SSL would inevitably introduce auxiliary noises to the representations under the target behavior, under which circumstance we propose intra-behavior SSL. In order to study the capability of intra-behavior SSL to relieve the impact of noises, we artificially add a certain proportion of noisy interactions into the auxiliary data (\ie, 10\%, 20\%, 30\%) and then compare the performance decline percentage of MBSSL with that of several representative SOTA baselines (\ie, CML, S-MBRec, GHCF). Figure \ref{noise} shows the results on Beibei and Taobao.

We can observe that i) It is obvious that adding noises into the auxiliary data reduces the performance of all the methods. Generally, CML is more sensitive to the interaction noises while MBSSL presents the least performance degradation. What's more, the degradation gap between MBSSL and each baseline is more apparent as the noise ratio increases. This suggests the capability of MBSSL to figure out informative graph patterns and to relieve the over-reliance on certain interactions.
ii) The performance decrease of GHCF is lower than that of CML and S-MBRec in most of the cases which indicates that the SSL paradigm in CML and S-MBRec would amplify the negative impact of noises. However, the intra-behavior in MBSSL aims to counteract the noises via consolidating the target self-supervised signals, which consequently makes MBSSL obtain a more stable performance in terms of different noise ratios.

% the representations under the target behavior encounter the risk of negative transfer resulted from inter-behavior SSL. To solve this issue, we propose intra-behavior SSL with an aim to enlarge the impact of supervision signals under the target behavior.
\subsection{Impact of Hybrid Strategies (RQ4)}
Given the heuristic observation that there exists an obvious discrepancy between auxiliary tasks and the target task in both gradient direction and magnitude, we propose to manipulate the gradients of auxiliary tasks in these two dimensions to balance the optimization. Besides, to equip our model with a good generalization ability, we only alter auxiliary gradients with larger magnitudes than the target gradient while keeping auxiliary gradients with small magnitudes unaltered. In this subsection, with an aim to verify the rationality of our gradient manipulation method proposed in Section \ref{sec:hyb}, we compare three different strategies as follows:
% $\bullet$ \textbf{Strategy A}: In each iteration, we only alter the direction of auxiliary gradients following ~\cite{yu2020gradient}. Note that we manipulate the directions of all the eligible gradients regardless of their magnitudes.\\
% $\bullet$ \textbf{Strategy B}: In each iteration, we only alter the magnitude of auxiliary gradients following ~\cite{he2022metabalance}, \ie, we reduce the magnitude of the auxiliary gradient if it is larger than that of the target gradient and vice versa.\\
% $\bullet$ \textbf{Strategy C}: In each iteration, we first alter the direction using Strategy A, and then we alter the magnitude based on Strategy B. Similarly, we conduct the manipulation on all the gradients regardless of their magnitudes.\\
% $\bullet$ \textbf{Ours}: In each iteration, our strategy alters the direction and magnitude of auxiliary gradients successively. Note that we only alter the auxiliary gradients with larger magnitudes than target gradients.\\

\begin{itemize}[leftmargin=*]
    \item \textbf{Strategy A}: In each iteration, we only alter the direction of auxiliary gradients following ~\cite{yu2020gradient}. We manipulate the directions of all the eligible gradients regardless of their magnitudes.
    \item \textbf{Strategy B}: In each iteration, we only alter the magnitude of auxiliary gradients following ~\cite{he2022metabalance}, \ie, we reduce the magnitude of the auxiliary gradient if it is larger than that of the target gradient and vice versa.
    \item \textbf{Strategy C}: In each iteration, we first alter the direction using Strategy A, and then we alter the magnitude based on Strategy B. Similarly, we conduct the manipulation on all the gradients regardless of their magnitudes.
    % \item \textbf{Ours}: In each iteration, our strategy alters the direction and magnitude of auxiliary gradients successively. Note that we only alter the auxiliary gradients with larger magnitudes than target gradients.
\end{itemize}

\begin{table}[ht]
% \scriptsize
\caption{The comparison results of different strategies.}
\setlength{\abovecaptionskip}{1cm}
\setlength{\belowcaptionskip}{-1cm}
\footnotesize
\label{tab:strategy}
\centering
\begin{tabular}{c|cc|cc}

% \toprule
\hline
Data & \multicolumn{2}{c|}{Beibei} & \multicolumn{2}{c}{Taobao} \\
Metric  & Recall@10 & NDCG@10 & Recall@10 & NDCG@10 \\
\hline
\hline
Strategy A & 0.2081 & 0.1155 & 0.0950 & 0.0543 \\
Strategy B & 0.2196 & 0.1247 & 0.0994 & 0.0567 \\
Strategy C & 0.2114 & 0.1217 & 0.0942 & 0.0543 \\
\hline
\hline
MBSSL & \textbf{0.2229}  & \textbf{0.1277} & \textbf{0.1027} & \textbf{0.0576} \\
\hline

\end{tabular}
\end{table}

\vspace{-0.3cm}
We show the comparison results on Beibei and Taobao in Table \ref{tab:strategy}, the results on the other datasets are similar. From the table, we find that i) Solely utilizing direction-based or magnitude-based methods obtains suboptimal performance, which corresponds to our observation that the SSL task yields great disparity with the target task in both direction and magnitude. ii) The performance gap between Strategy C and our model suggests that deliberately keeping the conflict between the target gradient and auxiliary gradients with small magnitudes could improve the generalization ability and ease the overfitting issue.

% \enlargethispage{2em}
% \vspace{-0.5cm}
\subsection{Hyperparameter Study (RQ5)}
In this subsection, we conduct extensive experiments to examine the effects of several key hyperparameters, which include the number of propagation layers $L$, the temperature $\tau$, the relax factor $r$ and the number of eliminated false negatives $N$.
Figure \ref{hyper} shows the estimated performance decrease/increase percentage \textit{w.r.t} a randomly selected datum point.
% Note that we keep other hyperparameters the same for each study on one hyperparameter.

\vspace{-0.15cm}
\subsubsection{Effect of Propagation Layer Numbers} From Figure \ref{hyper}, we observe that more embedding propagation layers yield better performance due to the strengthened capability of capturing high-hop signals. However, when the number of layers exceeds 4, the performance suffers from a huge decline, which is resulted from the over-smoothing effect.
\vspace{-0.1cm}
\subsubsection{Effect of Temperature}
The temperature is tuned carefully in \{0.1, 0.2, 0.4, 0.8\}. According to the curves shown in Figure \ref{hyper}(b) and \ref{hyper}(e), we find that the best selection of $\tau$ varies by dataset. However, either too small (\eg, 0.1 ) or too large (\eg, 0.8) is not appropriate
% \todo{This may due to the fact that a high temperature value can lead to so similar values of similarities between nodes that the SSL task is too hard, while a small temperature value can enlarge the differences of similarities between nodes to make the SSL task neglect some important nodes with small similarity scores.}
 which suggests that a large temperature value will undermine the ability to distinguish between negative samples. Conversely, a too small value would excessively exaggerate the effects of some negative samples while making others useless.

\vspace{-0.1cm}
\subsubsection{Effect of Relax Factor}
Since relax factor controls the magnitude proximity between auxiliary gradients and target gradient, its selection varies according to the task.
% \todo{A high value of relax factor will force the magnitude of auxiliary tasks to be similar to the one of the target task, which will lead to the neglect of the magnitude of different tasks. However, a small value of relax factor cannot make the gradient change significantly, which means the strategy does not work well.}
Therefore, a big relax factor is appropriate on Videos while Beibei prefers a smaller one, and the others perform best with a moderate relax factor value.
\vspace{-0.1cm}
\subsubsection{Effect of Eliminated False Negatives Numbers}
\begin{figure}[t]
  \setlength{\belowcaptionskip}{-0.6cm}
  \centering
    % \subfigure[Number of Layers  5L$]{
    %     \begin{minipage}[]{0.3\linewidth}
    %     \includegraphics[width=1.15\linewidth]{hyper_layer_recall.pdf}
    %     \end{minipage}
    % }
    % \subfigure[Temperature $\tau$]{
    %     \begin{minipage}[]{0.3\linewidth}
    %     \includegraphics[width=1.15\linewidth]{hyper_temp_recall.pdf}
    %     \end{minipage}
    % }
    % \subfigure[Relax Factor $r$]{
    %     \begin{minipage}[]{0.3\linewidth}
    %     \includegraphics[width=1.15\linewidth]{hyper_metar_recall.pdf}
    %     \end{minipage}
    % }
    \subfigure[Number of Layers $L$]{
        \begin{minipage}[]{0.3\linewidth}
        \includegraphics[width=1.15\linewidth]{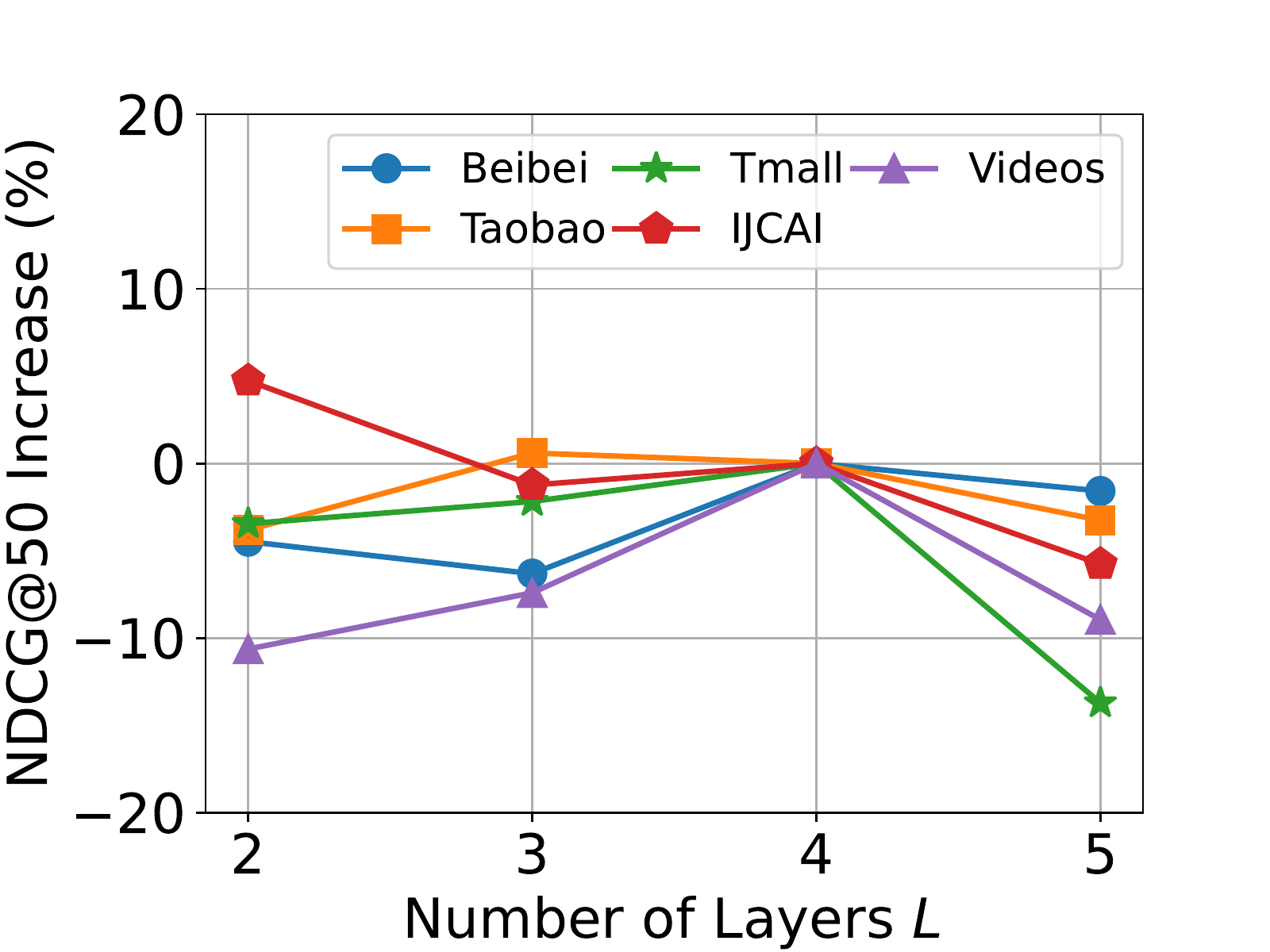}
        \end{minipage}
    }
    \subfigure[Temperature $\tau$]{
        \begin{minipage}[]{0.3\linewidth}
        \includegraphics[width=1.15\linewidth]{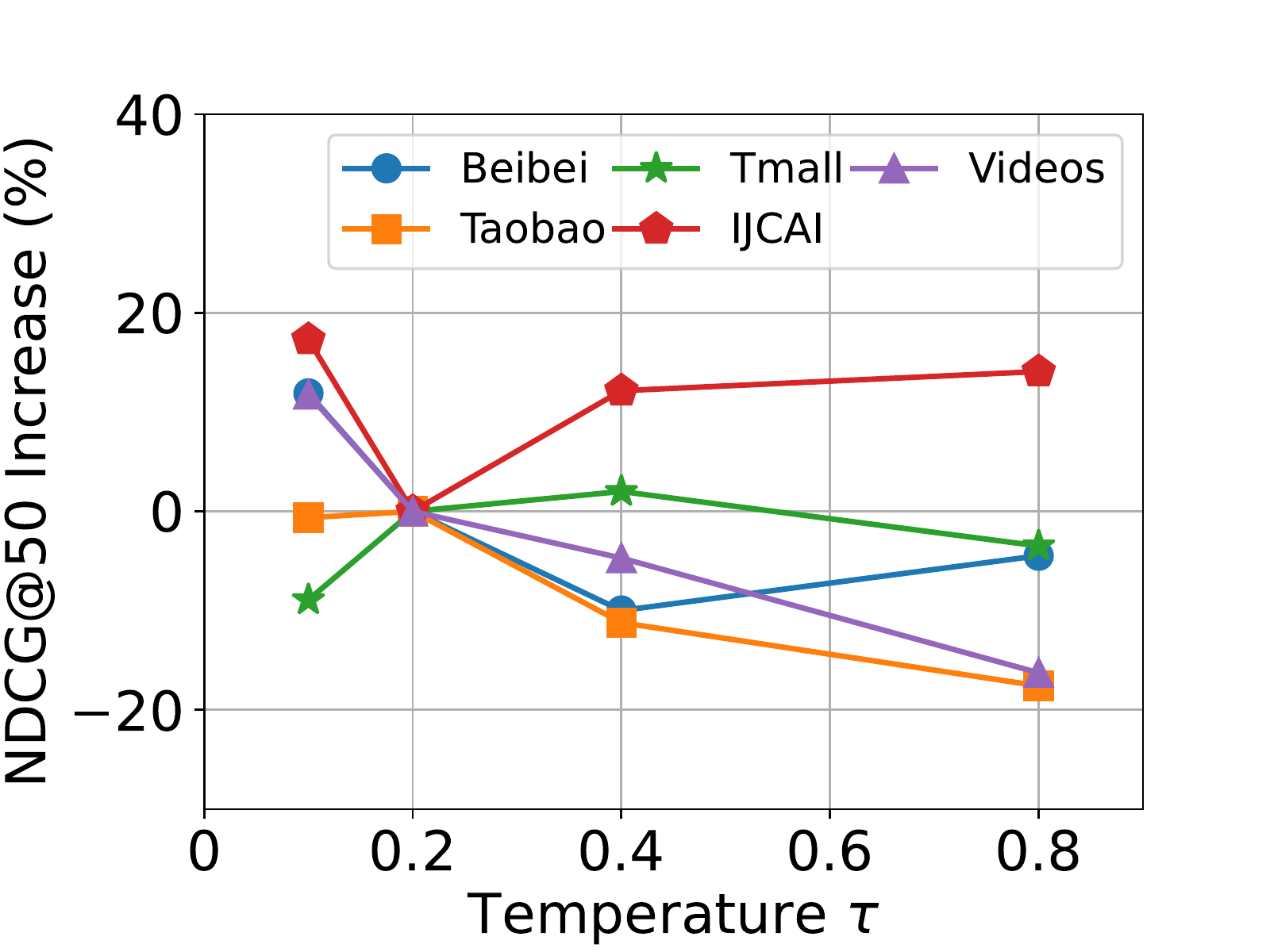}
        \end{minipage}
    }
    \subfigure[Relax Factor $r$]{
        \begin{minipage}[]{0.3\linewidth}
        \includegraphics[width=1.15\linewidth]{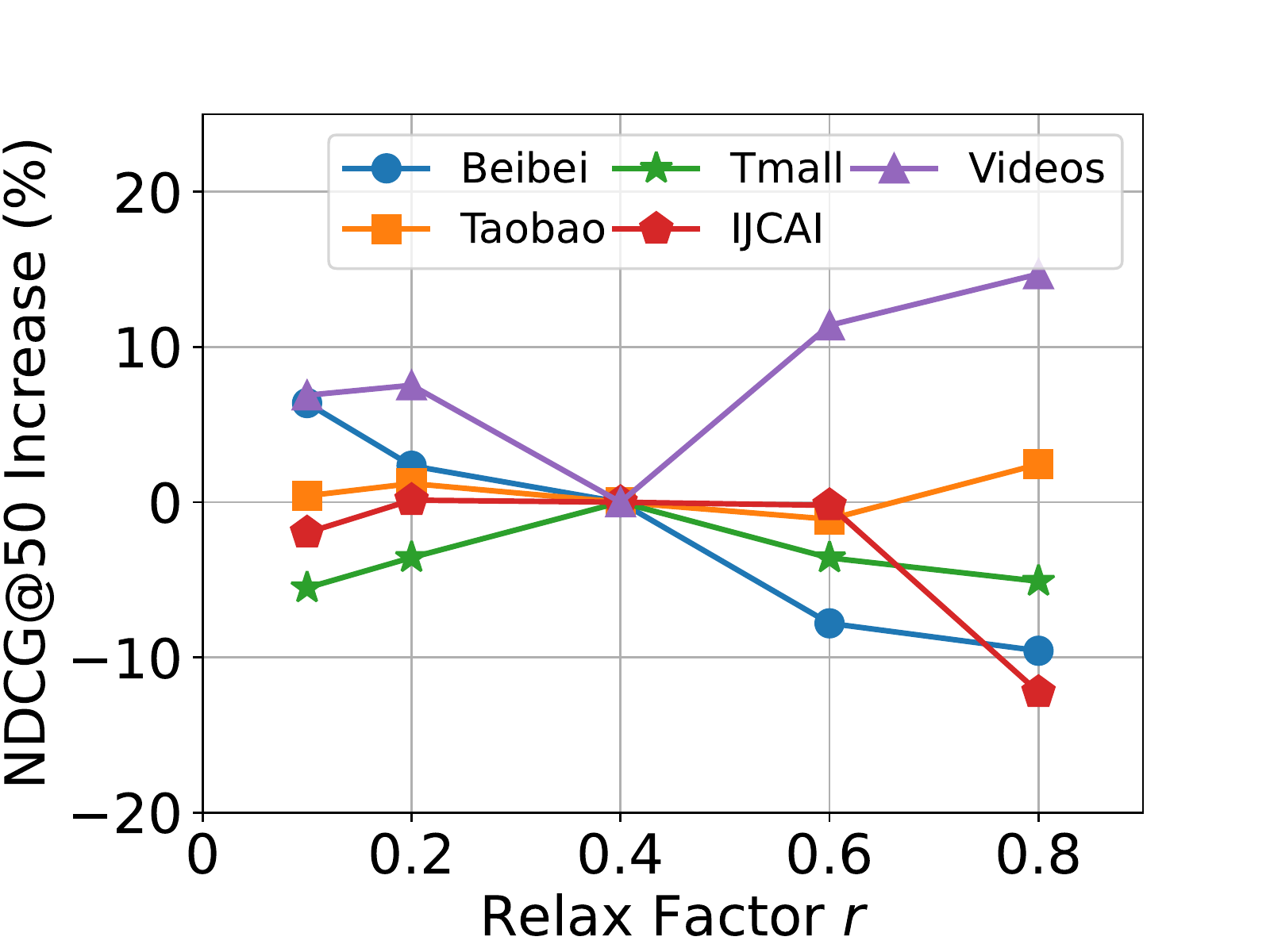}
        \end{minipage}
    }\vskip -12pt

    \subfigure[Eliminated Number $N$ on Beibei]{
        \begin{minipage}[]{0.45\linewidth}
         \centering
        \includegraphics[width=0.85\linewidth]{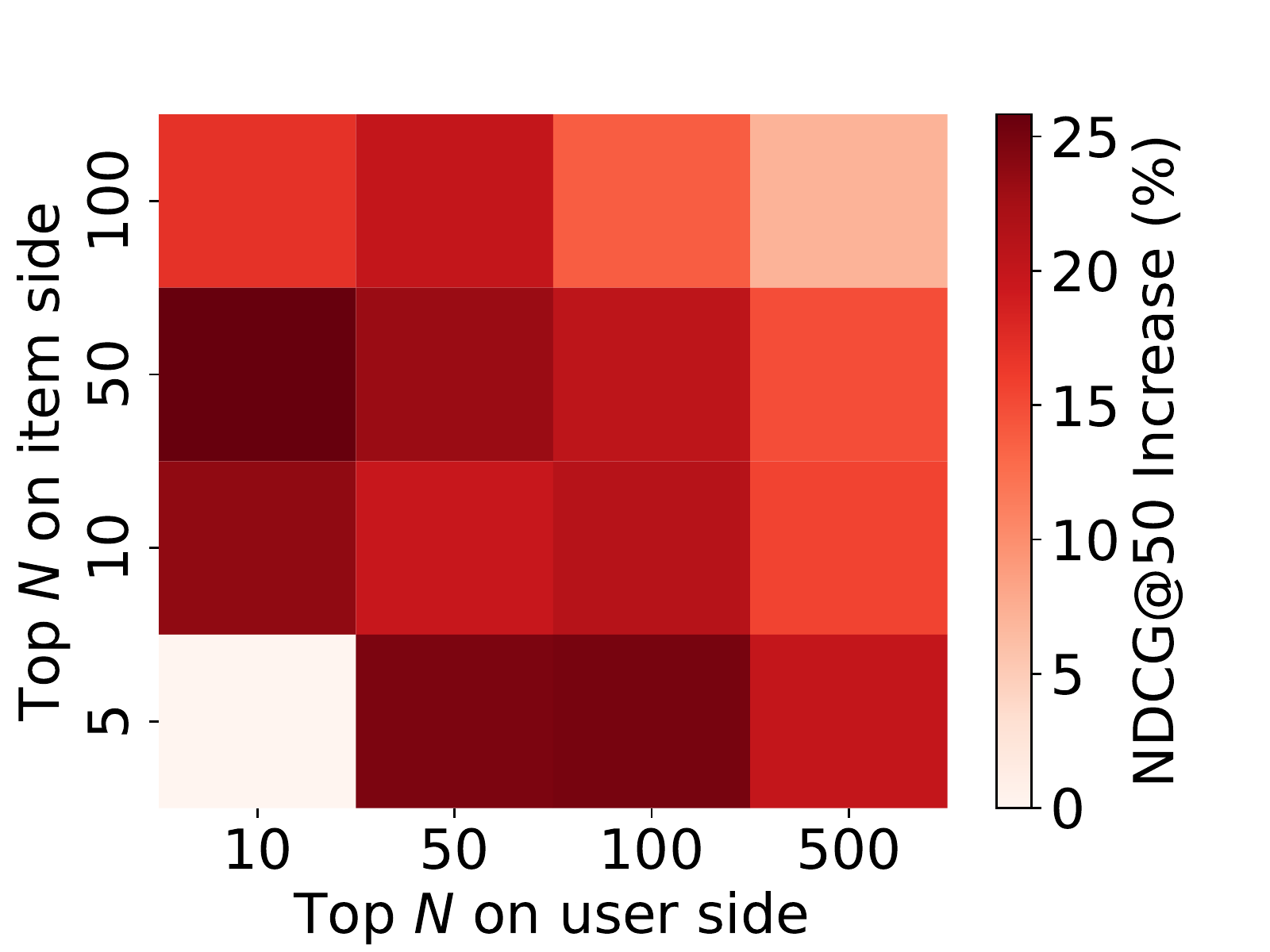}
        \end{minipage}
    }
    \subfigure[Eliminated Number $N$ on Taobao]{
        \begin{minipage}[]{0.45\linewidth}
        \centering
        \includegraphics[width=0.85\linewidth]{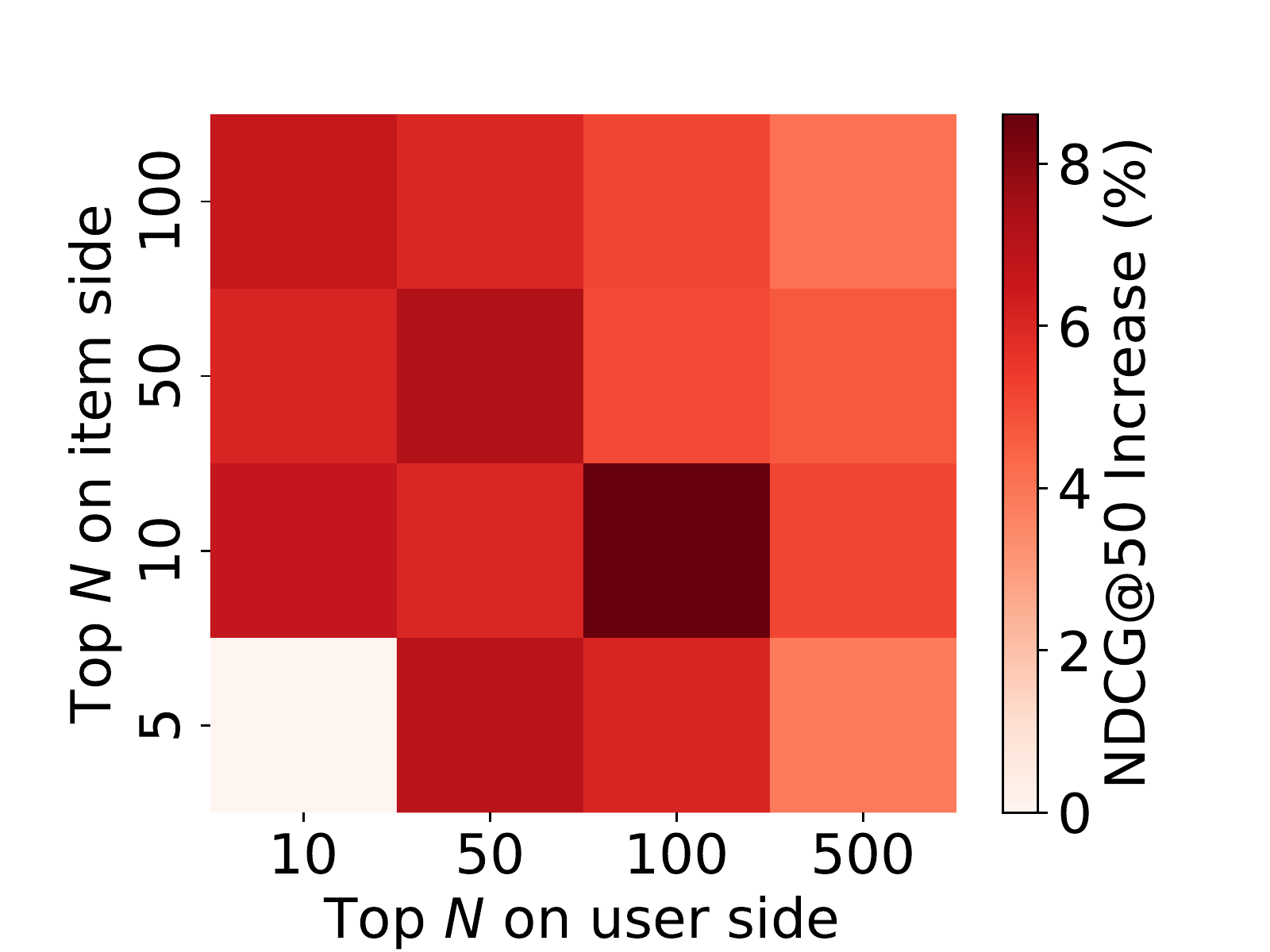}
        \end{minipage}
    }

  \caption{Hyperparameter study of MBSSL.}
  \label{hyper}
\end{figure}

Based on the average interaction numbers on each dataset, we determine the number of eliminated false negatives of user side and item side from \{10, 50, 100, 500\} and \{5, 10, 50, 100\} respectively. Figure \ref{hyper}(g) and \ref{hyper}(h)
show the results on Beibei and Taobao where darker color means better performance. We conclude that either too small or too big value of $N$ degrades the performance, which proves the necessity of sufficient negative samples and the effectiveness of the selective inter-behavior SSL.

\section{Related work}
\subsection{Graph-based Recommendation}

In terms of graph-based models for recommendation, recent years have witnessed the effectiveness of Graph Neural Networks (GNNs) due to the powerful capability on modeling high-order connectivity. For example, NGCF ~\cite{wang2019neural} is one collaborative architecture which decides the message propagation on both the graph structure and the affinity with the central node. LightGCN ~\cite{he2020lightgcn} distills a more concise and accurate GCN model for recommendation by omitting two burdensome operations, \ie, feature transformation and nonlinear activation. On top of methods focusing on model designs, another line of methods are dedicated to enriching the user-item bipartite graph via fusing various side information, ranging from social influences ~\cite{chen2019efficient, wu2019neural}, item-item relatedness ~\cite{wang2019kgat, wang2019knowledge, zhou2020interactive, wang2021learning}, to user and item attributes ~\cite{li2019fi, cen2019representation} .

\vspace{-0.2cm}
\subsection{Multi-behavior Recommendation}
% Behavior multiplicity is one important feature of user-item interactions, especially common in practical scenarios.
Recent studies have attempted to explore the multiplicity via various deep learning techniques. NMTR ~\cite{gao2019neural} extends the neural collaborative filtering (NCF ~\cite{he2017neural}) framework to multi-behavior settings, which performs a joint optimization on cascading prediction tasks. To avoid the sampling bias issue, EHCF ~\cite{chen2020efficient} efficiently correlates the prediction of each user behavior in a transfer way without negative sampling. Later on, researchers take advantages of GNNs to explore the high-hop user-item interactions. MBGCN ~\cite{jin2020multi} and GHCF ~\cite {chen2021graph} learn discriminative behavior representations using GCNs while MBGMN ~\cite{xia2021graph} utilizes graph meta network to capture interaction diversity and behavior heterogeneity. To alleviate data sparsity of the target behavior, CML~\cite{wei2022contrastive} and S-MBRec ~\cite{gu2022self} incorporate self-supervised learning into multi-behavior recommendation.

\vspace{-0.2cm}
\subsection{Self-supervised Learning on Graphs}
% Self-supervised learning (SSL) has emerged as a common paradigm to extract intrinsic data dependencies directly from unlabeled data.
% It initially achieved immense success in the domains of computer vision (CV) and natural language processing (NLP) ~\cite{chen2020simple, he2020momentum, fang2020cert, wu2020clear}. Early SSL methods design various prediction pretext tasks for representation learning, e.g., image transformation ~\cite{pathak2016context, zhang2016colorful, noroozi2016unsupervised} in CV or language modeling ~\cite{devlin2018bert, yang2019xlnet, giorgi2020declutr} in NLP, which drive model to learn the correlations among different data forms. Up to now, the contrastive learning framework has become the leading component in SSL. It aims to learn quality representations via augmenting the raw data and contrasting data samples in different augmented views.
% Inspired by the fascinating advances in CV and NLP domain, there have been attempts to explore SSL on graph-structured data.
Inspired by the advances in CV and NLP domain, self-supervised learning (SSL) on graphs has recently been explored.
InfoGraph~\cite{sun2019infograph} and DGI ~\cite{velickovic2019deep} learn node representations based on mutual information maximization between a node and the local structure while GRACE~\cite{zhu2020deep}, GCA~\cite{zhu2021graph} and GraphCL ~\cite{hafidi2020graphcl} conduct node-level same-scale contrast. When it comes to the recommendation scenario, recent studies have adopted SSL to achieve better recommendation performance. Yao et al. ~\cite{yao2021self} propose a two-tower DNN architecture with uniform feature masking and dropout for self-supervised item recommendation. Wu et al. ~\cite{wu2021self} further devise a unified SSL framework with three augmentation operators for graph-based recommendation.

% \vspace{-8pt}
% \enlargethispage{2em}
\vspace{-0.2cm}
\section{Conclusion}
In this work, we develop a novel self-supervised learning framework with an adaptive optimization method for enhancing multi-behavior recommendation. Our framework effectively captures the behavior semantics and correlations via a graph neural network incorporating the self-attention mechanism. To alleviate the data sparsity and noisy interactions issue, we contrast nodes via inter-behavior SSL and intra-behavior SSL respectively. In addition, we take the initial step to study the optimization imbalance of the SSL task and recommendation task, and design a hybrid manipulation method on gradients accordingly, which has been proved effective on five real-world datasets. In the future, we plan to design a more elaborated SSL by fully capturing the structural information based on the various interaction data.
% For verifying the rationality of our MBSSL, we compared it with various SOTA methods on several real-world datasets and
% performed experiments including the ablation study and hyperparameter study for a comprehensive illustration.

% \section*{Acknowledgments}

\begin{acks}
This work is supported in part by the National Natural Science Foundation of China (No.~61872207) and Kuaishou Inc.
Chaokun Wang is the corresponding author.
\end{acks}

\bibliographystyle{ACM-Reference-Format}
\bibliography{sample-base}
% \appendix

% \section{Appendix}








\end{document}